\documentclass[%
 reprint,
 amsmath,amssymb,
 aps,
]{revtex4-2}
\usepackage[version=3]{mhchem}
\usepackage{graphicx}% Include figure files
\usepackage{dcolumn}% Align table columns on decimal point
\usepackage{bm}% bold math
\begin{document}

\preprint{APS/123-QED}

\title{Luttinger compensated bipolarized magnetic semiconductor}% Force line breaks with \\
\author{Peng-Jie Guo$^{1,2}$}
\email{guopengjie@ruc.edu.cn}
\author{Huan-Cheng Yang$^{1,2}$}
\author{Xiao-Yao Hou$^{1,2}$}
\author{Ze-Feng Gao$^{1,2}$}
\author{Wei Ji$^{1,2}$}
\email{wji@ruc.edu.cn}
\author{Zhong-Yi Lu$^{1,2,3}$}
\email{zlu@ruc.edu.cn}

\affiliation{1. Department of Physics and Beijing Key Laboratory of Opto-electronic Functional Materials $\&$ Micro-nano Devices. Renmin University of China, Beijing 100872, China}
\affiliation{2. Key Laboratory of Quantum State Construction and Manipulation (Ministry of Education), Renmin University of China, Beijing 100872, China}
\affiliation{3. Hefei National Laboratory, Hefei 230088, China}

\date{\today}% It is always \today, today,

\begin{abstract}
Altermagnetic materials, with real-space antiferromagnetic arrangement and reciprocal-space anisotropic spin splitting, have attracted much attention. However, the spin splitting is small in most altermagnetic materials, which is a disadvantage to their application in electronic devices. In this study, based on symmetry analysis and the first-principles electronic structure calculations, we predict for the first time two Luttinger compensated bipolarized magnetic semiconductors Mn(CN)$_2$ and Co(CN)$_2$ with isotropic spin splitting as in the ferromagnetic materials. Our further analysis shows that the Luttinger compensated magnetism here depends not only on spin group symmetry, but also on the crystal field splitting and the number of d-orbital
electrons. In addition, the polarized charge density indicates that both Mn(CN)$_2$ and Co(CN)$_2$ have the quasi-symmetry $\rm{T\tau}$, resulting from the crystal field splitting and the number of $d$-orbital electrons. 
%We further propose a class of Luttinger compensated ferrimagnetic materials protected by quasi-symmetry, which share many characteristics with altermagnetic materials.
The Luttinger compensated magnetism not only has the zero total  magnetic moment as the antiferromagnetism, but also has the isotropic spin splitting as the ferromagnetism, thus our work not only provides theoretical guidance for searching Luttinger compensated magnetic materials with distinctive properties, but also provides a material basis for the application in spintronic devices.
\end{abstract}

\maketitle

%\tableofcontents

%\section{\label{sec:intro}Introduction }
{\it Introduction.} Ferromagnetic (FM) semiconductors with spintronic and transistor functionalities have strongly attracted experimental and theoretical interest due to their potential applications in next-generation electronic devices\cite{science-spintronics, NM-FMS, DMS-RM, science-2D, JJAP-FM-spintronics}. However, ferromagnetic materials are usually more likely to be metals than insulators. Although many ferromagnetic semiconductors have been proposed\cite{CrI3, CrBr3, CrGeTe3, CrSiTe3, NiBr2, PRR}, their magnetic transition temperatures are relatively low, which hinders their application in electronic devices. In addition, there are another two shortcomings. First, a key disadvantage of ferromagnetic semiconductors is the effect of stray fields due to non-zero net magnetic moments. Second, the slow reversal of ferromagnetic domain is not conducive to the improvement of device speed. Compared with ferromagnetic materials, antiferromagnetic materials are usually insulators very likely with high Neel temperatures, and may take the advantages of feasible miniaturization due to zero total magnetic moments and fast domain reversal speed. An interesting question is whether or not there are magnetic materials that simultaneously share the respective excellent features of ferromagnetic semiconductors and antiferromagnetic insulators. 

Recently, based on spin group theory, altermagnetism distincting from ferromagnetism and conventional antiferromagnetism has been proposed \cite{PRX-1, PRX-2}. Interestingly, altermagnetic materials may simultaneously share the respective excellent features of ferromagnetic materials and antiferromagnetic materials but without their disadvantages. Thus, altermagnetic materials show many novel physical properties \cite{altermagnetism-1, altermagnetism-2, altermagnetism-3, altermagnetism-4, PRX-1, PRX-2, PRX-3, QAH-npj2023, SST-PRL2021, SST-PRL2022, SST-PRL2022-2,SST-NE2022, GMR-PRX2022, GMR-2024, TMR-Shao2021, SC-AM, AHE-Sinova2022, AHE-RuO2-NE2022, AHE-MnTe-PRL2023, AHE-hou2023, MOE-Yao2021, CTHE-Yao2024, HighoT-liu2024, MCM-liu2023, LiFe2F6-guo2023, NiF3-qu2024, BWS, CVHE}. In altermagnetism, opposite spin lattices cannot be connected by space-inversion symmetry $\left\{C_2||I\right\}$ or fractional translation symmetry $\left\{C_2||\tau \right\}$, but by rotation or mirror symmetry $\left\{C_2||R\right\}$. Here, the symmetry operations at the left and right of the double vertical bar act only on the spin space and lattice space, respectively; the notation $C_2$ represents the $180^{\circ} $ rotation perpendicular to the spin direction; the notations $I, R$, and $\tau$ denote space inversion, rotation/mirror, and fractional translation operations, respectively. The spin symmetry $\left\{C_2||R\right\}$ guarantees zero total magnetic moment, while the breaking of the spin symmetry$\left\{C_2||I\right\}$ and $\left\{C_2||\tau \right\}$ leads to altermagnetic materials with anisotropic spin splitting, such as $d$-wave, $g$-wave and $i$-wave spin splitting. Different from ferromagnetic materials, the magnitude of spin splitting in altermagnetic materials depends on the magnitude of the difference between hoppings along different directions, resulting from the anisotropic crystal field \cite{BWS}. But, the crystal field anisotropy of magnetic atoms with opposite magnetic moments for most altermagnetic materials is not strong, so their spin splitting is usually small. Up to now, although more than 200 altermagnetic materials have been predicted\cite{Song-PRX, Liu-PRX, Gao-AI}, only very few altermagnetic materials have large spin splitting, which is not conducive to the application of altermagnetic materials in electronic devices. 

%there may be a  magnetic phase named Luttinger compensated magnet\cite{MazinPRX2022}

Inspired by the definition of altermagnetism, we consider such a magnetic phase in which the opposite spin lattices cannot be connected by any crystal symmetry, but the total magnetic moment in the unit cell is exactly zero. When a material is an insulator or half-metal, the spin magnetic moments per unit cell can have only integer value in Bohr magnetons by Luttinger's theorem\cite{Luttinger-I,Luttinger-II}. Consequently, if the total spin magnetic moment of the material is relatively small, it will be exactly zero. Further, if the orbital magnetic moments of the material is simultaneously completely quenched in the crystal field, its total magnetic moment will be zero. This new magnetic phase is named as Luttinger compensated magnetic phase\cite{MazinPRX2022}.  Different from altermagentic materials, the Luttinger compensated magnetic materials have isotropic spin polarization, which is similar to ferromagnetic materials and beneficial to improve the spin polarization of carriers. Thus, the Luttinger compensated magnetism may simultaneously also take the respective advantage of ferromagnetism and antiferromagnetism. Moreover, considering that the Luttinger compensation magnetism generally has an isotropic spin splitting and completely polarized carriers, it may have more advantages than altermagnetism in electronic device applications. Then, the most concerned issue is whether or not this novel magnetic phase can be realized in realistic materials.

In this study, we predict two Luttinger compensated bipolarized magnetic semiconductors Mn(CN)$_2$ and Co(CN)$_2$, whose opposite spin lattices cannot be connected by any crystal symmetry but with fully compensated total magnetic moments based on symmetry analysis and the first-principles electronic structure calculations. Moreover, the polarized charge density indicates that both Mn(CN)$_2$ and Co(CN)$_2$ have the quasi-symmetry $\rm{T\tau}$, resulting from the crystal field splitting and the number of $d$-orbital electrons. Finally, we propose a scheme to search for  the Luttinger compensated ferrimagnetic materials.

{\it Method.} Our electronic structure calculations employed the Vienna Ab initio Simulation Package (VASP)\cite{Vasp-1996} with the projector augmented wave (PAW) method\cite{PAW-1994}. The Perdew-Burke-Ernzerhof (PBE) exchange-correlation functional \cite{GGA-1996} and the GGA plus on-site repulsion $U$ method (GGA+$U$) were used in our calculations\cite{LDAU1, LDAU2}. The kinetic energy cutoff was set to be 600 eV for the expanding the wave functions into a plane-wave basis and the energy convergence criterion is $10^{-6}$ eV. The $\Gamma$-centered k-mesh was set as $12\times12\times12$. The crystal structures were fully relaxed until the force on each atom is less than 0.01 eV/\AA. The relaxed lattice parameters were 6.04 \AA and 5.75 \AA  for Co(CN)$_2$ and Mn(CN)$_2$, respectively. The Monte Carlo simulations based on the classical Heisenberg model were performed by using the open source project MCSOLVER\cite{Mcsolver}.

\begin{figure}[t]
\centering
\includegraphics[width=0.48\textwidth]{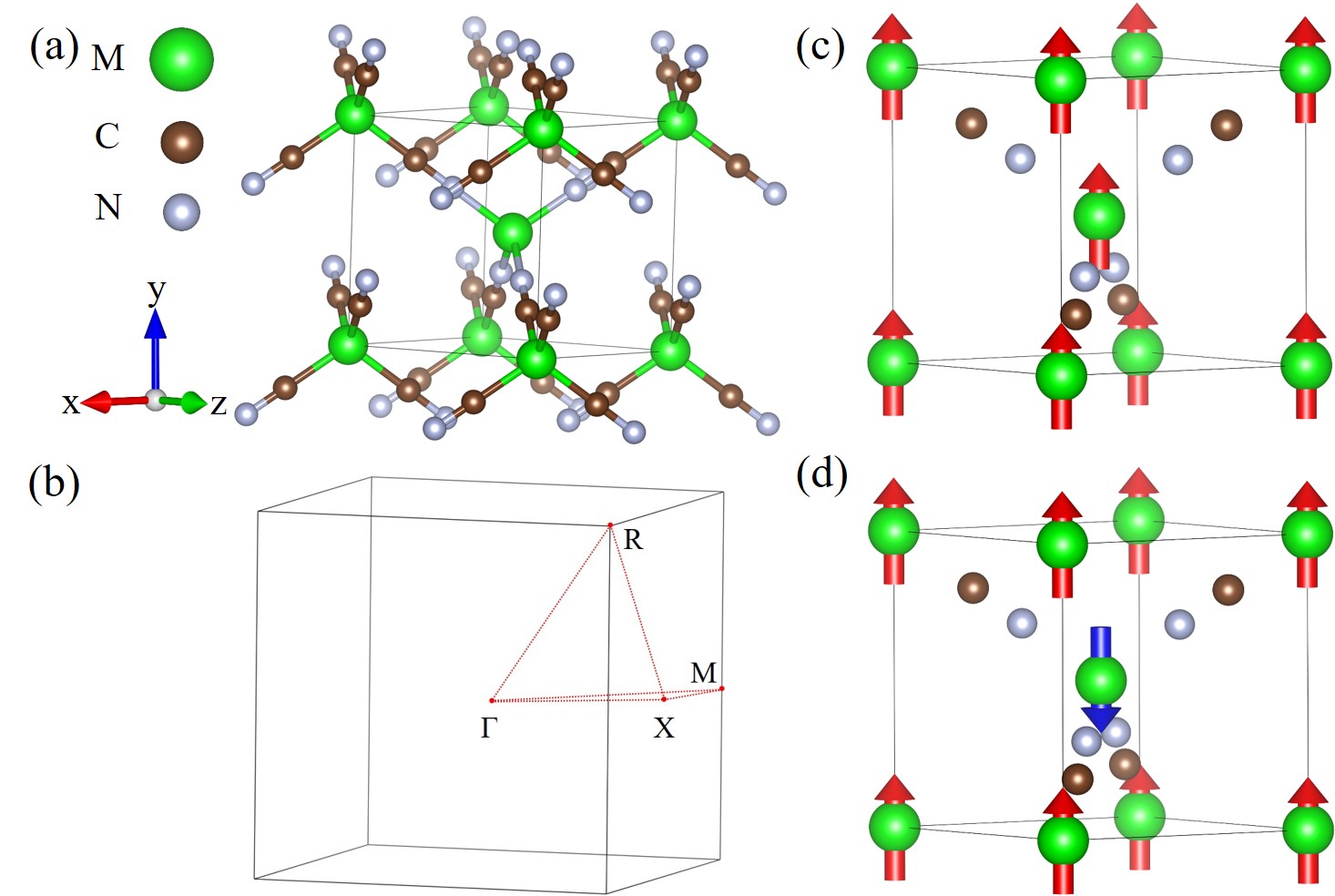}% Here is how to import EPS art
\caption{ The crystal structure and two typical magnetic structure of M(CN)$_2$. (a) and (b) are the crystal structure and Brillouin zone (BZ) of M(CN)$_2$. The red points represent high-symmetry points. Two typical collinear magnetic configurations of M(CN)$_2$: (c) the FM state, (d) the AFM state. The red and blue arrows represent the up and down spin moments, respectively.}
\label{fig:1}
\end{figure}
{\it Results and discussion.} For Luttinger compensated magnetic materials, the opposite spin sublattices need to have different crystal field environments to guarantee that they cannot be connected by any crystal symmetry. Here, we propose a cubic crystal structure M(M=Mn and Co)(CN)$_2$, in which two magnetic atoms M are surrounded by regular tetrahedra of C and N (Fig. 1(a)), respectively. Therefore, M(CN)$_2$ has P-43m (215) space group symmetry without space-inversion symmetry, and the corresponding point group symmetry is T$_d$. The Brillouin zone (BZ) of M(CN)$_2$ with high-symmetry lines and points is shown in Fig. 1 (b). To determine the magnetic structure of M(CN)$_2$, we consider two magnetic structures: ferromagnetic and antiferromagnetic, which are shown in Fig. 1 (c) and (d), respectively. We calculated the total energies of two magnetic structures forMn(CN)$_2$ and Co(CN)$_2$ with the variation of Hubbard interaction U, which are shown in Fig. 2 (a) and (c), respectively. From Fig. 2 (a) and (c), the antiferromagnetic state is always stable under different Hubbard interaction U for both Mn(CN)$_2$ and Co(CN)$_2$. On the other hand, since the angle between M-CN-M is of 180 degrees, M-M favors antiferromagnetic coupling according to GKA rules\cite{GKA}. Therefore, the results of electronic structure calculations and theoretical analysis  are consistent. 

%The antiferromagnetic state is magnetic ground state of both MnC$_2$N$_2$ and CoC$_2$N$_2$ (Fig. 1(d)). 

Due to the absence of space-inversion symmetry, M(CN)$_2$ has no $\left\{C_2||I\right\}$ symmetry. At the same time, M(CN)$_2$ has no $\left\{C_2||\tau \right\}$ symmetry due to the presence of C and N atoms in crystal cell. So both Mn(CN)$_2$ and Co(CN)$_2$ are not conventional collinear antiferromagnetic materials. Although the T$_d$ group has 24 symmetric operations including E, 8C$_3$, 3C$_2$, 6$\sigma_d$, 6S$_4$, none of them can connect the spin opposite magnetic atoms M in M(CN)$_2$. Therefore, both Mn(CN)$_2$ and Co(CN)$_2$ are not altermagnetic materials either, but may be Luttinger compensated magnetic materials. On the other hand, we also calculated the phonon spectra of Mn(CN)$_2$ and Co(CN)$_2$, which are shown in Fig2 (b) and (d), respectively, to check their dynamical stability. From Fig2 (b) and (d), neither Mn(CN)$_2$ nor Co(CN)$_2$ has imaginary frequencies in their phonon spectra. Thus, Mn(CN)$_2$ and Co(CN)$_2$ are dynamically stable.

\begin{figure}[t]
\centering
\includegraphics[width=0.48\textwidth]{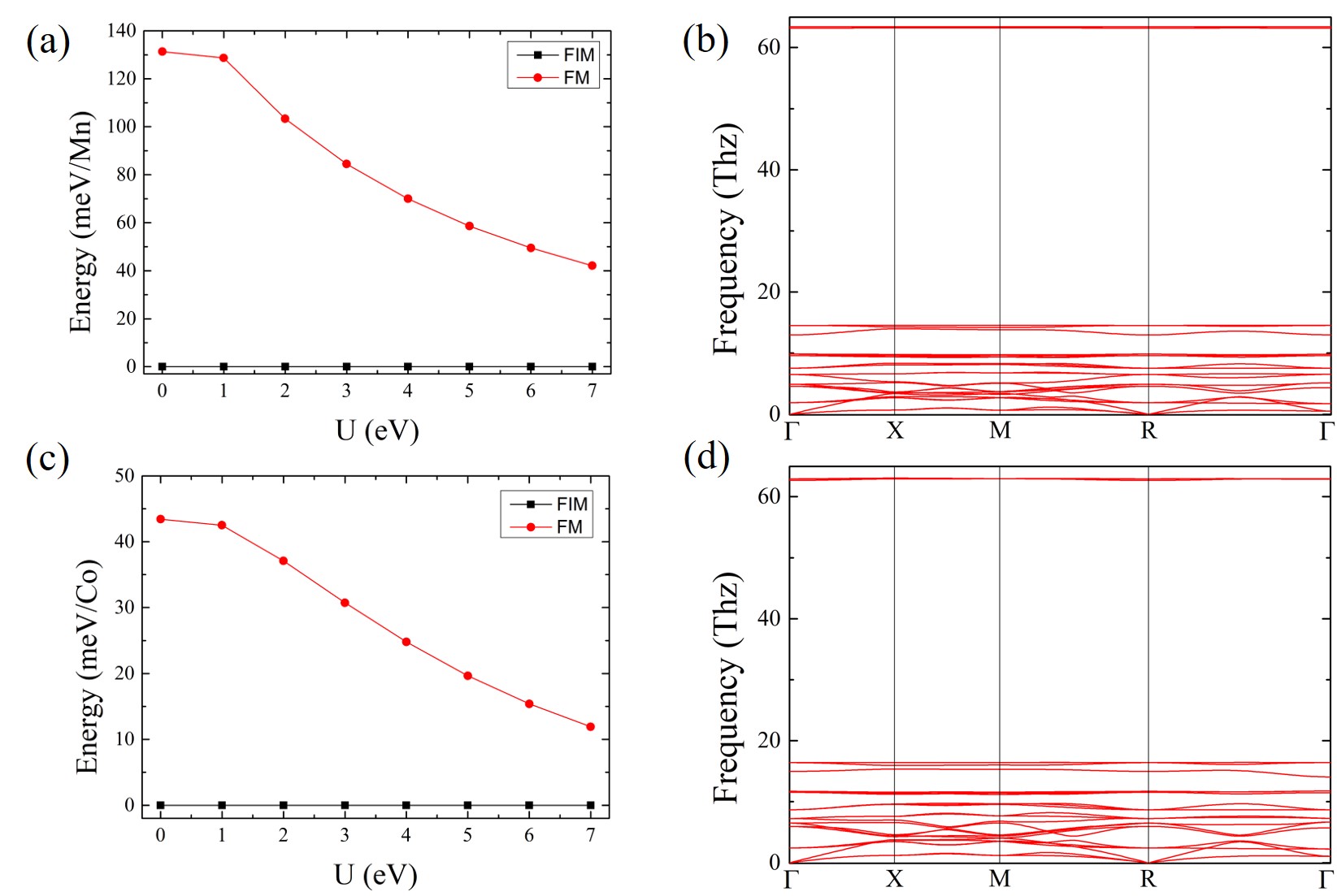}% Here is how to import EPS art
\caption{ Relative energy of different magnetic structures with variation of Hubbard interaction U and phonon spectra for M(CN)$_2$. (a) and (c) are relative energies of AFM state with respect to the FM state for Mn(CN)$_2$ and Co(CN)$_2$, respectively. (b) and (d) are the phonon spectrum of Mn(CN)$_2$ and Co(CN)$_2$, respectively. 
}
\label{fig:2}
\end{figure}

After verifying the structural stability and magnetic structure, we next study the electronic properties of Mn(CN)$_2$ and Co(CN)$_2$. Without spin-orbit coupling (SOC), both Mn(CN)$_2$ and Co(CN)$_2$ are semiconductors with bandgaps to be 3.89 eV and 2.4 eV as shown in Fig. 3(a) and (b), respectively. Different from altermagnetic materials, their spin-up and spin-down bands are completely split, which is the same as those of FM materials (Fig. 3(a) and (b)). This is because that atoms with opposite magnetic moments cannot be connected by any symmetry. But, the total magnetic moments of both Mn(CN)$_2$ and Co(CN)$_2$ are zero, which is the same as those of altermagnetic and conventional collinear antiferromagnetic materials. Interestingly, the highest occupied valence band and the lowest unoccupied conduction band have opposite spin polarizations for both Mn(CN)$_2$ and Co(CN)$_2$ (Fig. 3(a) and (b)), thus both Mn(CN)$_2$ and Co(CN)$_2$ are bipolarized magnetic semiconductors.

\begin{figure}[t]
\centering
\includegraphics[width=0.48\textwidth]{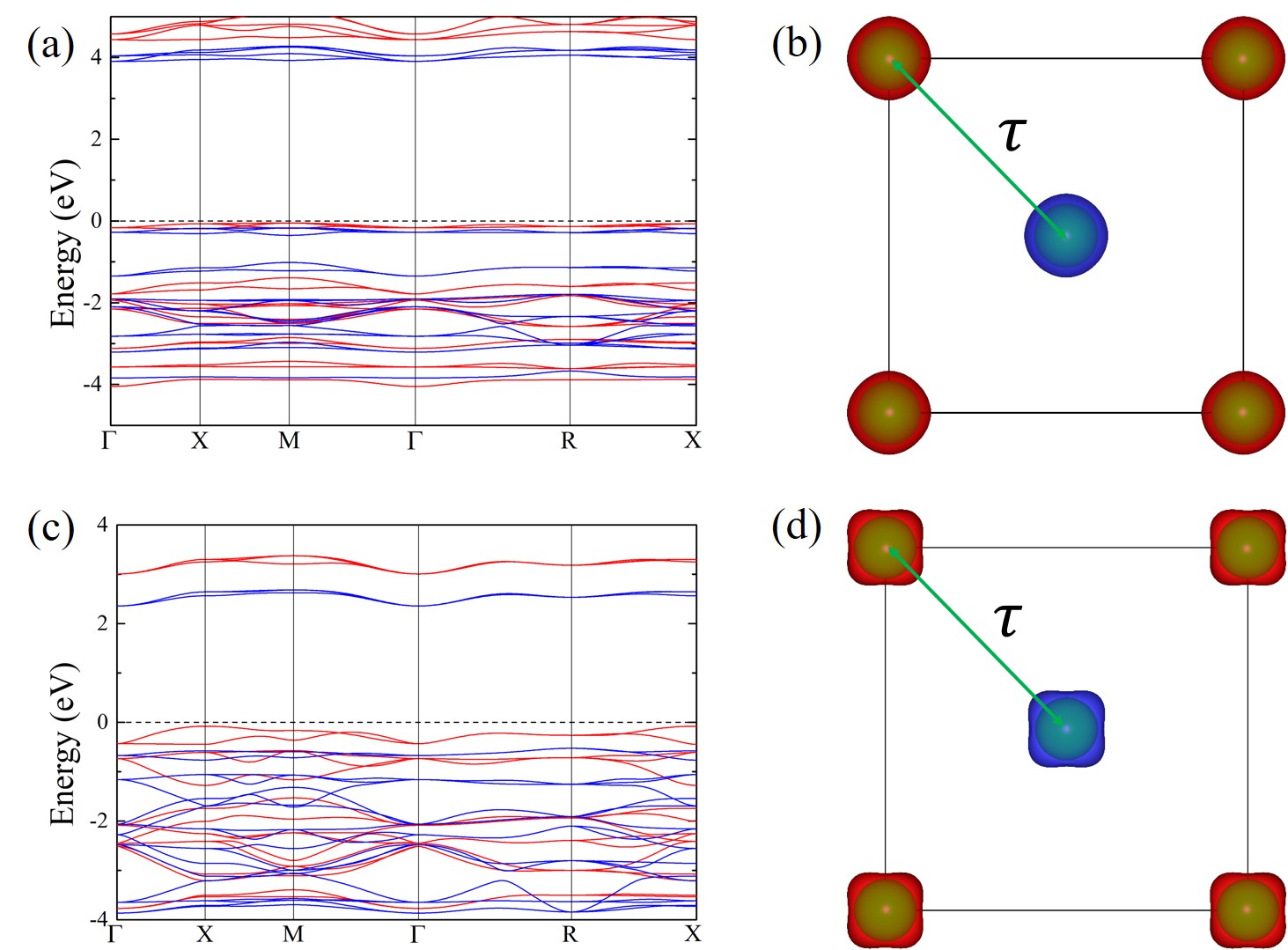}% Here is how to import EPS art
\caption{ The electronic band structure and polarized charge density of M(CN)$_2$. (a) and (c) are the electronic band structures of Mn(CN)$_2$ and Co(CN)$_2$ along the high-symmetry directions, respectively. The red and blue lines represent spin-up and spin-down channels, respectively. The polarized charge density of Mn(CN)$_2$ (b) and Co(CN)$_2$ (d). The red and blue three-dimensional surface represent respectively spin-up and spin-down polarized charge density. The $\tau$ represents (1/2, 1/2, 1/2) fractional translation. The electronic structure and polarized charge density are calculated with U=4 eV.
}
\label{fig:3}
\end{figure}

On the other hand, the magnetic transition temperature is very important for the application of magnetic materials. We estimated the magnetic transition temperatures T$_N$ of Mn(CN)$_2$ and Co(CN)$_2$ from the classical Monte Carlo simulations based on the three-dimensional lattice Heisenberg model. The Hamiltonian is shown as follow:
\begin{align}
H = J_1\sum_{i,j} S_i\cdot S_j
\label{Eq1}
\end{align}
where S$_i$ represents the spin of Co or Mn on site i and J$_1$ denotes the first-nearest neighbor exchange interaction parameters. The exchange interaction parameters were derived from mapping the energies of the FM and AFM magnetic configurations to the above model. The results are J$_1$S$^2$=3.10 and 8.75 meV for Co(CN)$_2$ and Mn(CN)$_2$, respectively. Then, the magnetic transition temperatures can be extracted from the peak of the specific-heat capacity based on the Monte Carlo simulations. As shown in Fig. 4 (a) and (b), the resultant magnetic transition temperatures for Mn(CN)$_2$ and Co(CN)$_2$ are 210 K and 75 K, respectively.

\begin{figure}[t]
\centering
\includegraphics[width=0.48\textwidth]{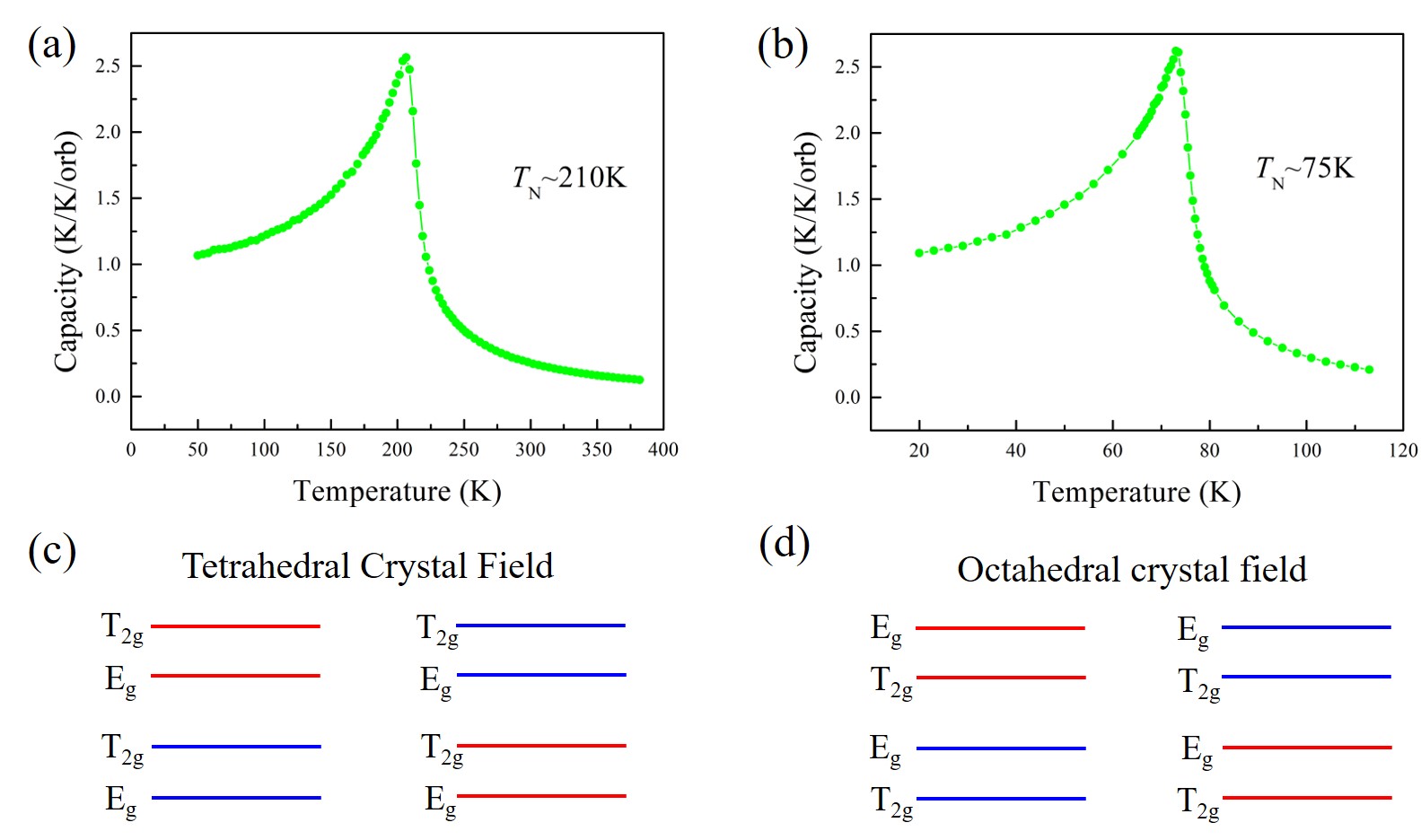}% Here is how to import EPS art
\caption{The magnetic transition temperatures and schematic diagram of crystal field splitting of M(CN)$_2$. (a) and (b) are evolution of specific heat capacity with temperature of Mn(CN)$_2$ and Co(CN)$_2$, respectively. (c) and (d) are the splitting schematics of $d$ orbitals in tetrahedral and octahedral crystal fields, respectively. The magnetic transition temperatures are calculated under correlation interaction U=4 eV.
}
\label{fig:4}
\end{figure}

An important question is why both Mn(CN)$_2$ and Co(CN)$_2$ have the exactly zero total magnetic moments. The question may be answered by the following three points. First, in the crystal of M(CN)$_2$, the Mn$^{+2}$ and Co$^{+2}$ ions are located in the regular tetrahedral crystal field. Therefore, the 3$d$ orbits will split into E$_g$ and T$_{2g}$ orbits and the E$_g$ orbits has lower energy than the T$_{2g}$ orbits as shown in Fig 4(c). The Mn$^{+2}$ and Co$^{+2}$ ions have 5 and 7 $d$ electrons, respectively. In the high-spin state, the orbital magnetic moments of both Mn$^{+2}$ and Co$^{+2}$ ions are completely quenched and the magnetic moments can only arise from spin moments. Second, both Mn(CN)$_2$ and Co(CN)$_2$ are insulators, which requires that they can have only an integer spin magnetic moment in Bohr magnetons per unit cell by Luttinger’s theorem\cite{Luttinger-I, Luttinger-II}. Finally, the opposite spin magnetic moments on Mn$^{+2}$ (Co$^{+2}$) in Mn(CN)$_2$ (Co(CN)$_2$) are almost equal in size, then the total spin magnetic moment in the unit cell should be very samll and it will then be exactly zero due to the constraint of  Luttinger’s theorem\cite{Luttinger-I, Luttinger-II}. On the other hand, we also calculated the polarized charge densities of Mn(CN)$_2$ and Co(CN)$_2$, which are shown in Fig.3(b) and (d), respectively. From Fig.3(b) and (d), the polarized charge density distribution of two Mn$^{+2}$ or Co$^{+2}$ with opposite magnetic moments are approximately equal, only the spins are opposite. Therefore, although both Mn(CN)$_2$ and Co(CN)$_2$ do not have crystal symmetry $ \rm {T\tau}$ they do have symmetry $ \rm {T\tau}$ at the charge density level. We call such a symmetry as quasi-symmetry. 

Considering another case, in Ti(CN)$_2$, 2 3-$d$ electrons of Ti$^{+2}$ ions fill exactly the low energy E$_g$ orbit, resulting in completely quenched orbital moments. Ti(CN)$_2$ is thus an insulator, and is also a Luttinger compensated magnetic material. It has also quasi-symmetry  $ \rm {T\tau}$, which is proved by our calculations. Unfortunately, the phonon spectrum of Ti(CN)$_2$ has imaginary frequencies, which makes Ti(CN)$_2$ unstable. If the E$_{g}$ orbital or T$_{2g}$ orbital is partially filled, the orbital moments of $d$ electrons will not be completely quenched and the materials should be metallic. By calculating V(CN)$_2$, Cr(CN)$_2$, Fe(CN)$_2$, and Ni(CN)$_2$, we find that the total magnetic moments of these magnetic materials is not exactly zero. For octahedral crystal field, the 3-$d$ orbits are also split into E$_g$ and T$_{2g}$ orbits, but the E$_g$ orbit is higher in energy than the T$_{2g}$ orbit as shown in Fig 4(d). Then, the $d$ electrons must be 3, 5 and 8 to guarantee the completely quenched orbital moments for high-spin states, in addition to satisfying the last two of the above three points, so that the material could be a Luttinger compensated magnet. Similarly, our analysis results can be also generalized to the materials containing 4$f$ atoms. In a word, the Luttinger compensated magnets depend not only on spin group symmetry, but also on the crystal field splitting and the number of d-orbital electrons. 

In the application of semiconductor, it is necessary to dope the semiconductor material. If an electron-type (hole-type) doping is applied to a compensated bipolarized magnetic semiconductor, this will produce fully polarized spin-up (spin-down) charge carriers. Meanwhile, the magnetic moment generated by doping is very small, which can avoid stray field effect. This is very important for applications in electronic devices. Although both Mn(CN)$_2$ and Co(CN)$_2$ are Luttinger compensated bipolarized magnetic semiconductor, their Neel temperatures are below room temperature. Therefore, in the future, we need to find Luttinger compensated magnetic materials with feasible properties, such as the transition temperature above the room temperature, the energy gap being in the semiconductor range, and the materials being nontoxicity.

In summary, based on symmetry analysis and the first-principles electronic structure calculations, we propose a class of Luttinger compensated bipolarized magnetic materials with quasi-symmetry $\rm{T\tau}$, represented by Mn(CN)$_2$ and Co(CN)$_2$. The full compensation (or exactly zero total magnetic moments) arise from: the completely quenched orbital magnetic moments, the insulativity of materials and the subsequent requirement that the spin magnetic moments per unit cell must be integer Bohr magnetons by Luttinger theorem. Considering that the Luttinger compensated magnetic materials take the advantages of both ferromagnetic spin splitting and antiferromagnetic zero total magnetic moments, our work not only enriches the classification of magnetic materials, but also provides new possibilities for applications in electronic devices.

%More importantly, If the spin-triplet excitons condense superfluid, the two spin superfluid phases with S=1 and -1 can be switched by electric field. 
%Due to the abundance of novel physical properties, \ce{LiFe2F6} will certainly attract a wide range of theoretical and experimental interest. 

\begin{acknowledgments}
We thank X.-H. Kong for valuable discussions. This work was financially supported by the National Natural Science Foundation of China (Grant No.12434009, No.12204533, No.62476278 and No.92477205), the National Key R$\&$D Program of China (Grant No.2024YFA1408601 and No.2023YFA1406500), the Fundamental Research Funds for the Central Universities, and the Research Funds of Renmin University of China (Grant No. 24XNKJ15 and No. 22XNKJ30). Computational resources have been provided by the Physical Laboratory of High Performance Computing at Renmin University of China.

P.-J.G. and H.-C. Y. contributed equally to this work.
\end{acknowledgments}

\nocite{*}

\bibliography{main}% Produces the bibliography via BibTeX.

\begin{thebibliography}{52}%
\makeatletter
\providecommand \@ifxundefined [1]{%
 \@ifx{#1\undefined}
}%
\providecommand \@ifnum [1]{%
 \ifnum #1\expandafter \@firstoftwo
 \else \expandafter \@secondoftwo
 \fi
}%
\providecommand \@ifx [1]{%
 \ifx #1\expandafter \@firstoftwo
 \else \expandafter \@secondoftwo
 \fi
}%
\providecommand \natexlab [1]{#1}%
\providecommand \enquote  [1]{``#1''}%
\providecommand \bibnamefont  [1]{#1}%
\providecommand \bibfnamefont [1]{#1}%
\providecommand \citenamefont [1]{#1}%
\providecommand \href@noop [0]{\@secondoftwo}%
\providecommand \href [0]{\begingroup \@sanitize@url \@href}%
\providecommand \@href[1]{\@@startlink{#1}\@@href}%
\providecommand \@@href[1]{\endgroup#1\@@endlink}%
\providecommand \@sanitize@url [0]{\catcode `\\12\catcode `\$12\catcode
  `\&12\catcode `\#12\catcode `\^12\catcode `\_12\catcode `\%12\relax}%
\providecommand \@@startlink[1]{}%
\providecommand \@@endlink[0]{}%
\providecommand \url  [0]{\begingroup\@sanitize@url \@url }%
\providecommand \@url [1]{\endgroup\@href {#1}{\urlprefix }}%
\providecommand \urlprefix  [0]{URL }%
\providecommand \Eprint [0]{\href }%
\providecommand \doibase [0]{https://doi.org/}%
\providecommand \selectlanguage [0]{\@gobble}%
\providecommand \bibinfo  [0]{\@secondoftwo}%
\providecommand \bibfield  [0]{\@secondoftwo}%
\providecommand \translation [1]{[#1]}%
\providecommand \BibitemOpen [0]{}%
\providecommand \bibitemStop [0]{}%
\providecommand \bibitemNoStop [0]{.\EOS\space}%
\providecommand \EOS [0]{\spacefactor3000\relax}%
\providecommand \BibitemShut  [1]{\csname bibitem#1\endcsname}%
\let\auto@bib@innerbib\@empty
%</preamble>
\bibitem [{\citenamefont {Wolf}\ \emph {et~al.}(2001)\citenamefont {Wolf},
  \citenamefont {Awschalom}, \citenamefont {Buhrman}, \citenamefont {Daughton},
  \citenamefont {von Molnar}, \citenamefont {Roukes}, \citenamefont
  {Chtchelkanova},\ and\ \citenamefont {Treger}}]{science-spintronics}%
  \BibitemOpen
  \bibfield  {author} {\bibinfo {author} {\bibfnamefont {S.~A.}\ \bibnamefont
  {Wolf}}, \bibinfo {author} {\bibfnamefont {D.~D.}\ \bibnamefont {Awschalom}},
  \bibinfo {author} {\bibfnamefont {R.~A.}\ \bibnamefont {Buhrman}}, \bibinfo
  {author} {\bibfnamefont {J.~M.}\ \bibnamefont {Daughton}}, \bibinfo {author}
  {\bibfnamefont {S.}~\bibnamefont {von Molnar}}, \bibinfo {author}
  {\bibfnamefont {M.~L.}\ \bibnamefont {Roukes}}, \bibinfo {author}
  {\bibfnamefont {A.~Y.}\ \bibnamefont {Chtchelkanova}},\ and\ \bibinfo
  {author} {\bibfnamefont {D.~M.}\ \bibnamefont {Treger}},\ }\bibfield  {title}
  {\bibinfo {title} {Spintronics: A spin-based electronics vision for the
  future},\ }\href {https://doi.org/10.1126/science.1065389} {\bibfield
  {journal} {\bibinfo  {journal} {Science}\ }\textbf {\bibinfo {volume}
  {294}},\ \bibinfo {pages} {1488} (\bibinfo {year} {2001})}\BibitemShut
  {NoStop}%
\bibitem [{\citenamefont {MacDonald}\ \emph {et~al.}(2005)\citenamefont
  {MacDonald}, \citenamefont {Schiffer},\ and\ \citenamefont
  {Samarth}}]{NM-FMS}%
  \BibitemOpen
  \bibfield  {author} {\bibinfo {author} {\bibfnamefont {A.~H.}\ \bibnamefont
  {MacDonald}}, \bibinfo {author} {\bibfnamefont {P.}~\bibnamefont
  {Schiffer}},\ and\ \bibinfo {author} {\bibfnamefont {N.}~\bibnamefont
  {Samarth}},\ }\bibfield  {title} {\bibinfo {title} {Ferromagnetic
  semiconductors: moving beyond (ga,mn)as},\ }\href
  {https://doi.org/10.1038/nmat1325} {\bibfield  {journal} {\bibinfo  {journal}
  {Nat. Mater.}\ }\textbf {\bibinfo {volume} {4}},\ \bibinfo {pages} {195}
  (\bibinfo {year} {2005})}\BibitemShut {NoStop}%
\bibitem [{\citenamefont {Dietl}\ and\ \citenamefont {Ohno}(2014)}]{DMS-RM}%
  \BibitemOpen
  \bibfield  {author} {\bibinfo {author} {\bibfnamefont {T.}~\bibnamefont
  {Dietl}}\ and\ \bibinfo {author} {\bibfnamefont {H.}~\bibnamefont {Ohno}},\
  }\bibfield  {title} {\bibinfo {title} {Dilute ferromagnetic semiconductors:
  Physics and spintronic structures},\ }\href
  {https://doi.org/10.1103/RevModPhys.86.187} {\bibfield  {journal} {\bibinfo
  {journal} {Rev. Mod. Phys.}\ }\textbf {\bibinfo {volume} {86}},\ \bibinfo
  {pages} {187} (\bibinfo {year} {2014})}\BibitemShut {NoStop}%
\bibitem [{\citenamefont {Gong}\ and\ \citenamefont
  {Zhang}(2019)}]{science-2D}%
  \BibitemOpen
  \bibfield  {author} {\bibinfo {author} {\bibfnamefont {C.}~\bibnamefont
  {Gong}}\ and\ \bibinfo {author} {\bibfnamefont {X.}~\bibnamefont {Zhang}},\
  }\bibfield  {title} {\bibinfo {title} {Two-dimensional magnetic crystals and
  emergent heterostructure devices},\ }\href
  {https://doi.org/10.1126/science.aav4450} {\bibfield  {journal} {\bibinfo
  {journal} {Science}\ }\textbf {\bibinfo {volume} {363}},\ \bibinfo {pages}
  {6428} (\bibinfo {year} {2019})}\BibitemShut {NoStop}%
\bibitem [{\citenamefont {Masaaki}(2020)}]{JJAP-FM-spintronics}%
  \BibitemOpen
  \bibfield  {author} {\bibinfo {author} {\bibfnamefont {T.}~\bibnamefont
  {Masaaki}},\ }\bibfield  {title} {\bibinfo {title} {Recent progress in
  ferromagnetic semiconductors and spintronics devices},\ }\href
  {https://doi.org/10.35848/1347-4065/abcadc} {\bibfield  {journal} {\bibinfo
  {journal} {Jpn. J. Appl. Phys.}\ }\textbf {\bibinfo {volume} {60}},\ \bibinfo
  {pages} {010101} (\bibinfo {year} {2020})}\BibitemShut {NoStop}%
\bibitem [{\citenamefont {Huang}\ \emph {et~al.}(2017)\citenamefont {Huang},
  \citenamefont {Clark}, \citenamefont {Navarro-Moratalla}, \citenamefont
  {Klein}, \citenamefont {Cheng}, \citenamefont {Seyler}, \citenamefont
  {Zhong}, \citenamefont {Schmidgall}, \citenamefont {McGuire}, \citenamefont
  {Cobden}, \citenamefont {Yao}, \citenamefont {Xiao}, \citenamefont
  {Jarillo-Herrero},\ and\ \citenamefont {Xu}}]{CrI3}%
  \BibitemOpen
  \bibfield  {author} {\bibinfo {author} {\bibfnamefont {B.}~\bibnamefont
  {Huang}}, \bibinfo {author} {\bibfnamefont {G.}~\bibnamefont {Clark}},
  \bibinfo {author} {\bibfnamefont {E.}~\bibnamefont {Navarro-Moratalla}},
  \bibinfo {author} {\bibfnamefont {D.~R.}\ \bibnamefont {Klein}}, \bibinfo
  {author} {\bibfnamefont {R.}~\bibnamefont {Cheng}}, \bibinfo {author}
  {\bibfnamefont {K.~L.}\ \bibnamefont {Seyler}}, \bibinfo {author}
  {\bibfnamefont {D.}~\bibnamefont {Zhong}}, \bibinfo {author} {\bibfnamefont
  {E.}~\bibnamefont {Schmidgall}}, \bibinfo {author} {\bibfnamefont {M.~A.}\
  \bibnamefont {McGuire}}, \bibinfo {author} {\bibfnamefont {D.~H.}\
  \bibnamefont {Cobden}}, \bibinfo {author} {\bibfnamefont {W.}~\bibnamefont
  {Yao}}, \bibinfo {author} {\bibfnamefont {D.}~\bibnamefont {Xiao}}, \bibinfo
  {author} {\bibfnamefont {P.}~\bibnamefont {Jarillo-Herrero}},\ and\ \bibinfo
  {author} {\bibfnamefont {X.~D.}\ \bibnamefont {Xu}},\ }\bibfield  {title}
  {\bibinfo {title} {Layer-dependent ferromagnetism in a van der waals crystal
  down to the monolayer limit},\ }\href {https://doi.org/10.1038/nature22391}
  {\bibfield  {journal} {\bibinfo  {journal} {Nature}\ }\textbf {\bibinfo
  {volume} {546}},\ \bibinfo {pages} {270} (\bibinfo {year}
  {2017})}\BibitemShut {NoStop}%
\bibitem [{\citenamefont {Zhang}\ \emph {et~al.}(2019)\citenamefont {Zhang},
  \citenamefont {Shang}, \citenamefont {Jiang}, \citenamefont {Rasmita},
  \citenamefont {Gao},\ and\ \citenamefont {Yu}}]{CrBr3}%
  \BibitemOpen
  \bibfield  {author} {\bibinfo {author} {\bibfnamefont {Z.~W.}\ \bibnamefont
  {Zhang}}, \bibinfo {author} {\bibfnamefont {J.~Z.}\ \bibnamefont {Shang}},
  \bibinfo {author} {\bibfnamefont {C.~Y.}\ \bibnamefont {Jiang}}, \bibinfo
  {author} {\bibfnamefont {A.}~\bibnamefont {Rasmita}}, \bibinfo {author}
  {\bibfnamefont {W.~B.}\ \bibnamefont {Gao}},\ and\ \bibinfo {author}
  {\bibfnamefont {T.}~\bibnamefont {Yu}},\ }\bibfield  {title} {\bibinfo
  {title} {Direct photoluminescence probing of ferromagnetism in monolayer
  two-dimensional crbr$_3$},\ }\href
  {https://doi.org/10.1021/acs.nanolett.9b00553} {\bibfield  {journal}
  {\bibinfo  {journal} {Nano Lett.}\ }\textbf {\bibinfo {volume} {19}},\
  \bibinfo {pages} {3138} (\bibinfo {year} {2019})}\BibitemShut {NoStop}%
\bibitem [{\citenamefont {Gong}\ \emph {et~al.}(2017)\citenamefont {Gong},
  \citenamefont {Li}, \citenamefont {Li}, \citenamefont {Ji}, \citenamefont
  {Stern}, \citenamefont {Xia}, \citenamefont {Cao}, \citenamefont {Bao},
  \citenamefont {Wang}, \citenamefont {Wang}, \citenamefont {Qiu},
  \citenamefont {Cava}, \citenamefont {Louie}, \citenamefont {Xia},\ and\
  \citenamefont {Zhang}}]{CrGeTe3}%
  \BibitemOpen
  \bibfield  {author} {\bibinfo {author} {\bibfnamefont {C.}~\bibnamefont
  {Gong}}, \bibinfo {author} {\bibfnamefont {L.}~\bibnamefont {Li}}, \bibinfo
  {author} {\bibfnamefont {Z.~L.}\ \bibnamefont {Li}}, \bibinfo {author}
  {\bibfnamefont {H.~W.}\ \bibnamefont {Ji}}, \bibinfo {author} {\bibfnamefont
  {A.}~\bibnamefont {Stern}}, \bibinfo {author} {\bibfnamefont
  {Y.}~\bibnamefont {Xia}}, \bibinfo {author} {\bibfnamefont {T.}~\bibnamefont
  {Cao}}, \bibinfo {author} {\bibfnamefont {W.}~\bibnamefont {Bao}}, \bibinfo
  {author} {\bibfnamefont {C.~Z.}\ \bibnamefont {Wang}}, \bibinfo {author}
  {\bibfnamefont {Y.}~\bibnamefont {Wang}}, \bibinfo {author} {\bibfnamefont
  {Z.~Q.}\ \bibnamefont {Qiu}}, \bibinfo {author} {\bibfnamefont {R.~J.}\
  \bibnamefont {Cava}}, \bibinfo {author} {\bibfnamefont {S.~G.}\ \bibnamefont
  {Louie}}, \bibinfo {author} {\bibfnamefont {J.}~\bibnamefont {Xia}},\ and\
  \bibinfo {author} {\bibfnamefont {X.}~\bibnamefont {Zhang}},\ }\bibfield
  {title} {\bibinfo {title} {Discovery of intrinsic ferromagnetism in
  two-dimensional van der waals crystals},\ }\href
  {https://doi.org/10.1038/nature22060} {\bibfield  {journal} {\bibinfo
  {journal} {Nature}\ }\textbf {\bibinfo {volume} {546}},\ \bibinfo {pages}
  {265} (\bibinfo {year} {2017})}\BibitemShut {NoStop}%
\bibitem [{\citenamefont {Li}\ and\ \citenamefont {Yang}(2014)}]{CrSiTe3}%
  \BibitemOpen
  \bibfield  {author} {\bibinfo {author} {\bibfnamefont {X.~X.}\ \bibnamefont
  {Li}}\ and\ \bibinfo {author} {\bibfnamefont {J.~L.}\ \bibnamefont {Yang}},\
  }\bibfield  {title} {\bibinfo {title} {Crxte$_3$ (x = si, ge) nanosheets: two
  dimensional intrinsic ferromagnetic semiconductors},\ }\href
  {https://doi.org/10.1039/C4TC01193G} {\bibfield  {journal} {\bibinfo
  {journal} {J. Mater. Chem. C}\ }\textbf {\bibinfo {volume} {2}},\ \bibinfo
  {pages} {7071} (\bibinfo {year} {2014})}\BibitemShut {NoStop}%
\bibitem [{\citenamefont {Kulish}\ and\ \citenamefont {Huang}(2017)}]{NiBr2}%
  \BibitemOpen
  \bibfield  {author} {\bibinfo {author} {\bibfnamefont {V.~V.}\ \bibnamefont
  {Kulish}}\ and\ \bibinfo {author} {\bibfnamefont {W.}~\bibnamefont {Huang}},\
  }\bibfield  {title} {\bibinfo {title} {Single-layer metal halides mx2 (x =
  cl, br, i): stability and tunable magnetism from first principles and monte
  carlo simulations},\ }\href {https://doi.org/10.1039/C7TC02664A} {\bibfield
  {journal} {\bibinfo  {journal} {J. Mater. Chem. C}\ }\textbf {\bibinfo
  {volume} {5}},\ \bibinfo {pages} {8734} (\bibinfo {year} {2017})}\BibitemShut
  {NoStop}%
\bibitem [{\citenamefont {You}\ \emph {et~al.}(2020)\citenamefont {You},
  \citenamefont {Zhang}, \citenamefont {Dong}, \citenamefont {Gu},\ and\
  \citenamefont {Su}}]{PRR}%
  \BibitemOpen
  \bibfield  {author} {\bibinfo {author} {\bibfnamefont {J.~Y.}\ \bibnamefont
  {You}}, \bibinfo {author} {\bibfnamefont {Z.}~\bibnamefont {Zhang}}, \bibinfo
  {author} {\bibfnamefont {X.~J.}\ \bibnamefont {Dong}}, \bibinfo {author}
  {\bibfnamefont {B.}~\bibnamefont {Gu}},\ and\ \bibinfo {author}
  {\bibfnamefont {G.}~\bibnamefont {Su}},\ }\bibfield  {title} {\bibinfo
  {title} {Two-dimensional magnetic semiconductors with room curie
  temperatures},\ }\href {https://doi.org/10.1103/PhysRevResearch.2.013002}
  {\bibfield  {journal} {\bibinfo  {journal} {Phys. Rev. Res.}\ }\textbf
  {\bibinfo {volume} {2}},\ \bibinfo {pages} {013002} (\bibinfo {year}
  {2020})}\BibitemShut {NoStop}%
\bibitem [{\citenamefont {$\check{\rm{S}}$mejkal}\ \emph
  {et~al.}(2022{\natexlab{a}})\citenamefont {$\check{\rm{S}}$mejkal},
  \citenamefont {Sinova},\ and\ \citenamefont {Jungwirth}}]{PRX-1}%
  \BibitemOpen
  \bibfield  {author} {\bibinfo {author} {\bibfnamefont {L.}~\bibnamefont
  {$\check{\rm{S}}$mejkal}}, \bibinfo {author} {\bibfnamefont {J.}~\bibnamefont
  {Sinova}},\ and\ \bibinfo {author} {\bibfnamefont {T.}~\bibnamefont
  {Jungwirth}},\ }\bibfield  {title} {\bibinfo {title} {{Emerging Research
  Landscape of Altermagnetism}},\ }\href
  {https://doi.org/10.1103/PhysRevX.12.040501} {\bibfield  {journal} {\bibinfo
  {journal} {Phys. Rev. X}\ }\textbf {\bibinfo {volume} {12}},\ \bibinfo
  {pages} {040501} (\bibinfo {year} {2022}{\natexlab{a}})}\BibitemShut
  {NoStop}%
\bibitem [{\citenamefont {$\check{\rm{S}}$mejkal}\ \emph
  {et~al.}(2022{\natexlab{b}})\citenamefont {$\check{\rm{S}}$mejkal},
  \citenamefont {Sinova},\ and\ \citenamefont {Jungwirth}}]{PRX-2}%
  \BibitemOpen
  \bibfield  {author} {\bibinfo {author} {\bibfnamefont {L.}~\bibnamefont
  {$\check{\rm{S}}$mejkal}}, \bibinfo {author} {\bibfnamefont {J.}~\bibnamefont
  {Sinova}},\ and\ \bibinfo {author} {\bibfnamefont {T.}~\bibnamefont
  {Jungwirth}},\ }\bibfield  {title} {\bibinfo {title} {{Beyond Conventional
  Ferromagnetism and Antiferromagnetism: A Phase with Nonrelativistic Spin and
  Crystal Rotation Symmetry}},\ }\href
  {https://doi.org/10.1103/PhysRevX.12.031042} {\bibfield  {journal} {\bibinfo
  {journal} {Phys. Rev. X}\ }\textbf {\bibinfo {volume} {12}},\ \bibinfo
  {pages} {031042} (\bibinfo {year} {2022}{\natexlab{b}})}\BibitemShut
  {NoStop}%
\bibitem [{\citenamefont {Hayami}\ \emph {et~al.}(2019)\citenamefont {Hayami},
  \citenamefont {Yanagi},\ and\ \citenamefont {Kusunose}}]{altermagnetism-1}%
  \BibitemOpen
  \bibfield  {author} {\bibinfo {author} {\bibfnamefont {S.}~\bibnamefont
  {Hayami}}, \bibinfo {author} {\bibfnamefont {Y.}~\bibnamefont {Yanagi}},\
  and\ \bibinfo {author} {\bibfnamefont {H.}~\bibnamefont {Kusunose}},\
  }\bibfield  {title} {\bibinfo {title} {{Momentum-Dependent Spin Splitting by
  Collinear Antiferromagnetic Ordering}},\ }\href
  {https://doi.org/10.7566/JPSJ.88.123702} {\bibfield  {journal} {\bibinfo
  {journal} {J. Phys. Soc. Jpn.}\ }\textbf {\bibinfo {volume} {88}},\ \bibinfo
  {pages} {123702} (\bibinfo {year} {2019})}\BibitemShut {NoStop}%
\bibitem [{\citenamefont {$\check{\rm{S}}$mejkal}\ \emph
  {et~al.}(2020)\citenamefont {$\check{\rm{S}}$mejkal}, \citenamefont
  {Gonz$\acute{\rm{a}}$lez-Hern$\acute{\rm{a}}$ndez}, \citenamefont
  {Jungwirth},\ and\ \citenamefont {Sinova}}]{altermagnetism-2}%
  \BibitemOpen
  \bibfield  {author} {\bibinfo {author} {\bibfnamefont {L.}~\bibnamefont
  {$\check{\rm{S}}$mejkal}}, \bibinfo {author} {\bibfnamefont {R.}~\bibnamefont
  {Gonz$\acute{\rm{a}}$lez-Hern$\acute{\rm{a}}$ndez}}, \bibinfo {author}
  {\bibfnamefont {T.}~\bibnamefont {Jungwirth}},\ and\ \bibinfo {author}
  {\bibfnamefont {J.}~\bibnamefont {Sinova}},\ }\bibfield  {title} {\bibinfo
  {title} {{Crystal time-reversal symmetry breaking and spontaneous Hall effect
  in collinear antiferromagnets}},\ }\href
  {https://doi.org/10.1126/sciadv.aaz8809} {\bibfield  {journal} {\bibinfo
  {journal} {Sci. Adv.}\ }\textbf {\bibinfo {volume} {6}},\ \bibinfo {pages}
  {eaaz8809} (\bibinfo {year} {2020})}\BibitemShut {NoStop}%
\bibitem [{\citenamefont {Yuan}\ \emph {et~al.}(2020)\citenamefont {Yuan},
  \citenamefont {Wang}, \citenamefont {Luo}, \citenamefont {Rashba},\ and\
  \citenamefont {Zunger}}]{altermagnetism-3}%
  \BibitemOpen
  \bibfield  {author} {\bibinfo {author} {\bibfnamefont {L.-D.}\ \bibnamefont
  {Yuan}}, \bibinfo {author} {\bibfnamefont {Z.}~\bibnamefont {Wang}}, \bibinfo
  {author} {\bibfnamefont {J.-W.}\ \bibnamefont {Luo}}, \bibinfo {author}
  {\bibfnamefont {E.~I.}\ \bibnamefont {Rashba}},\ and\ \bibinfo {author}
  {\bibfnamefont {A.}~\bibnamefont {Zunger}},\ }\bibfield  {title} {\bibinfo
  {title} {{Giant momentum-dependent spin splitting in centrosymmetric low-$Z$
  antiferromagnets}},\ }\href {https://doi.org/10.1103/PhysRevB.102.014422}
  {\bibfield  {journal} {\bibinfo  {journal} {Phys. Rev. B}\ }\textbf {\bibinfo
  {volume} {102}},\ \bibinfo {pages} {014422} (\bibinfo {year}
  {2020})}\BibitemShut {NoStop}%
\bibitem [{\citenamefont {Mazin}\ \emph {et~al.}(2021)\citenamefont {Mazin},
  \citenamefont {Koepernik}, \citenamefont {Johannes}, \citenamefont
  {Gonz$\acute{\rm{a}}$lez-Hern$\acute{\rm{a}}$ndez},\ and\ \citenamefont
  {$\check{\rm{S}}$mejkal}}]{altermagnetism-4}%
  \BibitemOpen
  \bibfield  {author} {\bibinfo {author} {\bibfnamefont {I.~I.}\ \bibnamefont
  {Mazin}}, \bibinfo {author} {\bibfnamefont {K.}~\bibnamefont {Koepernik}},
  \bibinfo {author} {\bibfnamefont {M.~D.}\ \bibnamefont {Johannes}}, \bibinfo
  {author} {\bibfnamefont {R.}~\bibnamefont
  {Gonz$\acute{\rm{a}}$lez-Hern$\acute{\rm{a}}$ndez}},\ and\ \bibinfo {author}
  {\bibfnamefont {L.}~\bibnamefont {$\check{\rm{S}}$mejkal}},\ }\bibfield
  {title} {\bibinfo {title} {{Prediction of unconventional magnetism in doped
  FeSb$_2$}},\ }\href {https://doi.org/10.1073/pnas.2108924118} {\bibfield
  {journal} {\bibinfo  {journal} {Proc. Natl. Acad. Sci. U.S.A.}\ }\textbf
  {\bibinfo {volume} {118}},\ \bibinfo {pages} {e2108924118} (\bibinfo {year}
  {2021})}\BibitemShut {NoStop}%
\bibitem [{\citenamefont {Mazin}(2022{\natexlab{a}})}]{PRX-3}%
  \BibitemOpen
  \bibfield  {author} {\bibinfo {author} {\bibfnamefont {I.}~\bibnamefont
  {Mazin}},\ }\bibfield  {title} {\bibinfo {title} {Editorial:
  Altermagnetism---a new punch line of fundamental magnetism},\ }\href
  {https://doi.org/10.1103/PhysRevX.12.040002} {\bibfield  {journal} {\bibinfo
  {journal} {Phys. Rev. X}\ }\textbf {\bibinfo {volume} {12}},\ \bibinfo
  {pages} {040002} (\bibinfo {year} {2022}{\natexlab{a}})}\BibitemShut
  {NoStop}%
\bibitem [{\citenamefont {Guo}\ \emph {et~al.}(2023{\natexlab{a}})\citenamefont
  {Guo}, \citenamefont {Liu},\ and\ \citenamefont {Lu}}]{QAH-npj2023}%
  \BibitemOpen
  \bibfield  {author} {\bibinfo {author} {\bibfnamefont {P.-J.}\ \bibnamefont
  {Guo}}, \bibinfo {author} {\bibfnamefont {Z.-X.}\ \bibnamefont {Liu}},\ and\
  \bibinfo {author} {\bibfnamefont {Z.-Y.}\ \bibnamefont {Lu}},\ }\bibfield
  {title} {\bibinfo {title} {{Quantum anomalous hall effect in collinear
  antiferromagnetism}},\ }\href {https://doi.org/10.1038/s41524-023-01025-4}
  {\bibfield  {journal} {\bibinfo  {journal} {npj Comput. Mater.}\ }\textbf
  {\bibinfo {volume} {9}},\ \bibinfo {pages} {70} (\bibinfo {year}
  {2023}{\natexlab{a}})}\BibitemShut {NoStop}%
\bibitem [{\citenamefont {Gonz\'alez-Hern\'andez}\ \emph
  {et~al.}(2021)\citenamefont {Gonz\'alez-Hern\'andez}, \citenamefont
  {$\check{\rm{S}}$mejkal}, \citenamefont {V\'yborn\'y}, \citenamefont
  {Yahagi}, \citenamefont {Sinova}, \citenamefont {Jungwirth},\ and\
  \citenamefont {$\check{\rm{Z}}$elezn\'y}}]{SST-PRL2021}%
  \BibitemOpen
  \bibfield  {author} {\bibinfo {author} {\bibfnamefont {R.}~\bibnamefont
  {Gonz\'alez-Hern\'andez}}, \bibinfo {author} {\bibfnamefont {L.}~\bibnamefont
  {$\check{\rm{S}}$mejkal}}, \bibinfo {author} {\bibfnamefont {K.}~\bibnamefont
  {V\'yborn\'y}}, \bibinfo {author} {\bibfnamefont {Y.}~\bibnamefont {Yahagi}},
  \bibinfo {author} {\bibfnamefont {J.}~\bibnamefont {Sinova}}, \bibinfo
  {author} {\bibfnamefont {T.~c.~v.}\ \bibnamefont {Jungwirth}},\ and\ \bibinfo
  {author} {\bibfnamefont {J.}~\bibnamefont {$\check{\rm{Z}}$elezn\'y}},\
  }\bibfield  {title} {\bibinfo {title} {{Efficient Electrical Spin Splitter
  Based on Nonrelativistic Collinear Antiferromagnetism}},\ }\href
  {https://doi.org/10.1103/PhysRevLett.126.127701} {\bibfield  {journal}
  {\bibinfo  {journal} {Phys. Rev. Lett.}\ }\textbf {\bibinfo {volume} {126}},\
  \bibinfo {pages} {127701} (\bibinfo {year} {2021})}\BibitemShut {NoStop}%
\bibitem [{\citenamefont {Bai}\ \emph {et~al.}(2022)\citenamefont {Bai},
  \citenamefont {Han}, \citenamefont {Feng}, \citenamefont {Zhou},
  \citenamefont {Su}, \citenamefont {Wang}, \citenamefont {Liao}, \citenamefont
  {Zhu}, \citenamefont {Chen}, \citenamefont {Pan}, \citenamefont {Fan},\ and\
  \citenamefont {Song}}]{SST-PRL2022}%
  \BibitemOpen
  \bibfield  {author} {\bibinfo {author} {\bibfnamefont {H.}~\bibnamefont
  {Bai}}, \bibinfo {author} {\bibfnamefont {L.}~\bibnamefont {Han}}, \bibinfo
  {author} {\bibfnamefont {X.~Y.}\ \bibnamefont {Feng}}, \bibinfo {author}
  {\bibfnamefont {Y.~J.}\ \bibnamefont {Zhou}}, \bibinfo {author}
  {\bibfnamefont {R.~X.}\ \bibnamefont {Su}}, \bibinfo {author} {\bibfnamefont
  {Q.}~\bibnamefont {Wang}}, \bibinfo {author} {\bibfnamefont {L.~Y.}\
  \bibnamefont {Liao}}, \bibinfo {author} {\bibfnamefont {W.~X.}\ \bibnamefont
  {Zhu}}, \bibinfo {author} {\bibfnamefont {X.~Z.}\ \bibnamefont {Chen}},
  \bibinfo {author} {\bibfnamefont {F.}~\bibnamefont {Pan}}, \bibinfo {author}
  {\bibfnamefont {X.~L.}\ \bibnamefont {Fan}},\ and\ \bibinfo {author}
  {\bibfnamefont {C.}~\bibnamefont {Song}},\ }\bibfield  {title} {\bibinfo
  {title} {{Observation of Spin Splitting Torque in a Collinear Antiferromagnet
  ${\mathrm{RuO}}_{2}$}},\ }\href
  {https://doi.org/10.1103/PhysRevLett.128.197202} {\bibfield  {journal}
  {\bibinfo  {journal} {Phys. Rev. Lett.}\ }\textbf {\bibinfo {volume} {128}},\
  \bibinfo {pages} {197202} (\bibinfo {year} {2022})}\BibitemShut {NoStop}%
\bibitem [{\citenamefont {Karube}\ \emph {et~al.}(2022)\citenamefont {Karube},
  \citenamefont {Tanaka}, \citenamefont {Sugawara}, \citenamefont {Kadoguchi},
  \citenamefont {Kohda},\ and\ \citenamefont {Nitta}}]{SST-PRL2022-2}%
  \BibitemOpen
  \bibfield  {author} {\bibinfo {author} {\bibfnamefont {S.}~\bibnamefont
  {Karube}}, \bibinfo {author} {\bibfnamefont {T.}~\bibnamefont {Tanaka}},
  \bibinfo {author} {\bibfnamefont {D.}~\bibnamefont {Sugawara}}, \bibinfo
  {author} {\bibfnamefont {N.}~\bibnamefont {Kadoguchi}}, \bibinfo {author}
  {\bibfnamefont {M.}~\bibnamefont {Kohda}},\ and\ \bibinfo {author}
  {\bibfnamefont {J.}~\bibnamefont {Nitta}},\ }\bibfield  {title} {\bibinfo
  {title} {{Observation of Spin-Splitter Torque in Collinear Antiferromagnetic
  ${\mathrm{RuO}}_{2}$}},\ }\href
  {https://doi.org/10.1103/PhysRevLett.129.137201} {\bibfield  {journal}
  {\bibinfo  {journal} {Phys. Rev. Lett.}\ }\textbf {\bibinfo {volume} {129}},\
  \bibinfo {pages} {137201} (\bibinfo {year} {2022})}\BibitemShut {NoStop}%
\bibitem [{\citenamefont {Bose}\ \emph {et~al.}(2022)\citenamefont {Bose},
  \citenamefont {Schreiber}, \citenamefont {Jain}, \citenamefont {Shao},
  \citenamefont {Nair}, \citenamefont {Sun}, \citenamefont {Zhang},
  \citenamefont {Muller}, \citenamefont {Tsymbal}, \citenamefont {Schlom},\
  and\ \citenamefont {Ralph}}]{SST-NE2022}%
  \BibitemOpen
  \bibfield  {author} {\bibinfo {author} {\bibfnamefont {A.}~\bibnamefont
  {Bose}}, \bibinfo {author} {\bibfnamefont {N.~J.}\ \bibnamefont {Schreiber}},
  \bibinfo {author} {\bibfnamefont {R.}~\bibnamefont {Jain}}, \bibinfo {author}
  {\bibfnamefont {D.-F.}\ \bibnamefont {Shao}}, \bibinfo {author}
  {\bibfnamefont {H.~P.}\ \bibnamefont {Nair}}, \bibinfo {author}
  {\bibfnamefont {J.}~\bibnamefont {Sun}}, \bibinfo {author} {\bibfnamefont
  {X.~S.}\ \bibnamefont {Zhang}}, \bibinfo {author} {\bibfnamefont {D.~A.}\
  \bibnamefont {Muller}}, \bibinfo {author} {\bibfnamefont {E.~Y.}\
  \bibnamefont {Tsymbal}}, \bibinfo {author} {\bibfnamefont {D.~G.}\
  \bibnamefont {Schlom}},\ and\ \bibinfo {author} {\bibfnamefont {D.~C.}\
  \bibnamefont {Ralph}},\ }\bibfield  {title} {\bibinfo {title} {Tilted spin
  current generated by the collinear antiferromagnet ruthenium dioxide},\
  }\href {https://doi.org/10.1038/s41928-022-00744-8} {\bibfield  {journal}
  {\bibinfo  {journal} {Nat. Electron.}\ }\textbf {\bibinfo {volume} {5}},\
  \bibinfo {pages} {267} (\bibinfo {year} {2022})}\BibitemShut {NoStop}%
\bibitem [{\citenamefont {$\check{\rm{S}}$mejkal}\ \emph
  {et~al.}(2022{\natexlab{c}})\citenamefont {$\check{\rm{S}}$mejkal},
  \citenamefont {Hellenes}, \citenamefont {Gonz\'alez-Hern\'andez},
  \citenamefont {Sinova},\ and\ \citenamefont {Jungwirth}}]{GMR-PRX2022}%
  \BibitemOpen
  \bibfield  {author} {\bibinfo {author} {\bibfnamefont {L.}~\bibnamefont
  {$\check{\rm{S}}$mejkal}}, \bibinfo {author} {\bibfnamefont {A.~B.}\
  \bibnamefont {Hellenes}}, \bibinfo {author} {\bibfnamefont {R.}~\bibnamefont
  {Gonz\'alez-Hern\'andez}}, \bibinfo {author} {\bibfnamefont {J.}~\bibnamefont
  {Sinova}},\ and\ \bibinfo {author} {\bibfnamefont {T.}~\bibnamefont
  {Jungwirth}},\ }\bibfield  {title} {\bibinfo {title} {{Giant and Tunneling
  Magnetoresistance in Unconventional Collinear Antiferromagnets with
  Nonrelativistic Spin-Momentum Coupling}},\ }\href
  {https://doi.org/10.1103/PhysRevX.12.011028} {\bibfield  {journal} {\bibinfo
  {journal} {Phys. Rev. X}\ }\textbf {\bibinfo {volume} {12}},\ \bibinfo
  {pages} {011028} (\bibinfo {year} {2022}{\natexlab{c}})}\BibitemShut
  {NoStop}%
\bibitem [{\citenamefont {Zhang}\ \emph {et~al.}(2024)\citenamefont {Zhang},
  \citenamefont {Cui}, \citenamefont {Li}, \citenamefont {Duan}, \citenamefont
  {Li}, \citenamefont {Yu},\ and\ \citenamefont {Yao}}]{GMR-2024}%
  \BibitemOpen
  \bibfield  {author} {\bibinfo {author} {\bibfnamefont {R.-W.}\ \bibnamefont
  {Zhang}}, \bibinfo {author} {\bibfnamefont {C.}~\bibnamefont {Cui}}, \bibinfo
  {author} {\bibfnamefont {R.}~\bibnamefont {Li}}, \bibinfo {author}
  {\bibfnamefont {J.}~\bibnamefont {Duan}}, \bibinfo {author} {\bibfnamefont
  {L.}~\bibnamefont {Li}}, \bibinfo {author} {\bibfnamefont {Z.-M.}\
  \bibnamefont {Yu}},\ and\ \bibinfo {author} {\bibfnamefont {Y.}~\bibnamefont
  {Yao}},\ }\bibfield  {title} {\bibinfo {title} {Predictable gate-field
  control of spin in altermagnets with spin-layer coupling},\ }\href
  {https://doi.org/10.1103/PhysRevLett.133.056401} {\bibfield  {journal}
  {\bibinfo  {journal} {Phys. Rev. Lett.}\ }\textbf {\bibinfo {volume} {133}},\
  \bibinfo {pages} {056401} (\bibinfo {year} {2024})}\BibitemShut {NoStop}%
\bibitem [{\citenamefont {Shao}\ \emph {et~al.}(2021)\citenamefont {Shao},
  \citenamefont {Zhang}, \citenamefont {Li}, \citenamefont {Eom},\ and\
  \citenamefont {Tsymbal}}]{TMR-Shao2021}%
  \BibitemOpen
  \bibfield  {author} {\bibinfo {author} {\bibfnamefont {D.-F.}\ \bibnamefont
  {Shao}}, \bibinfo {author} {\bibfnamefont {S.-H.}\ \bibnamefont {Zhang}},
  \bibinfo {author} {\bibfnamefont {M.}~\bibnamefont {Li}}, \bibinfo {author}
  {\bibfnamefont {C.-B.}\ \bibnamefont {Eom}},\ and\ \bibinfo {author}
  {\bibfnamefont {E.}~\bibnamefont {Tsymbal}},\ }\bibfield  {title} {\bibinfo
  {title} {Spin-neutral currents for spintronics},\ }\href
  {https://doi.org/10.1038/s41467-021-26915-3} {\bibfield  {journal} {\bibinfo
  {journal} {Nat. Commun.}\ }\textbf {\bibinfo {volume} {12}},\ \bibinfo
  {pages} {7061} (\bibinfo {year} {2021})}\BibitemShut {NoStop}%
\bibitem [{\citenamefont {Zhu}\ \emph {et~al.}(2023)\citenamefont {Zhu},
  \citenamefont {Zhuang}, \citenamefont {Wu},\ and\ \citenamefont
  {Yan}}]{SC-AM}%
  \BibitemOpen
  \bibfield  {author} {\bibinfo {author} {\bibfnamefont {D.}~\bibnamefont
  {Zhu}}, \bibinfo {author} {\bibfnamefont {Z.-Y.}\ \bibnamefont {Zhuang}},
  \bibinfo {author} {\bibfnamefont {Z.}~\bibnamefont {Wu}},\ and\ \bibinfo
  {author} {\bibfnamefont {Z.}~\bibnamefont {Yan}},\ }\bibfield  {title}
  {\bibinfo {title} {Topological superconductivity in two-dimensional
  altermagnetic metals},\ }\href {https://doi.org/10.1103/PhysRevB.108.184505}
  {\bibfield  {journal} {\bibinfo  {journal} {Phys. Rev. B}\ }\textbf {\bibinfo
  {volume} {108}},\ \bibinfo {pages} {184505} (\bibinfo {year}
  {2023})}\BibitemShut {NoStop}%
\bibitem [{\citenamefont {$\check{\rm{S}}$mejkal}\ \emph
  {et~al.}(2022{\natexlab{d}})\citenamefont {$\check{\rm{S}}$mejkal},
  \citenamefont {MacDonald}, \citenamefont {Sinova}, \citenamefont
  {Nakatsuji},\ and\ \citenamefont {Jungwirth}}]{AHE-Sinova2022}%
  \BibitemOpen
  \bibfield  {author} {\bibinfo {author} {\bibfnamefont {L.}~\bibnamefont
  {$\check{\rm{S}}$mejkal}}, \bibinfo {author} {\bibfnamefont {A.~H.}\
  \bibnamefont {MacDonald}}, \bibinfo {author} {\bibfnamefont {J.}~\bibnamefont
  {Sinova}}, \bibinfo {author} {\bibfnamefont {S.}~\bibnamefont {Nakatsuji}},\
  and\ \bibinfo {author} {\bibfnamefont {T.}~\bibnamefont {Jungwirth}},\
  }\bibfield  {title} {\bibinfo {title} {{Anomalous Hall antiferromagnets}},\
  }\href {https://doi.org/10.1038/s41578-022-00430-3} {\bibfield  {journal}
  {\bibinfo  {journal} {Nat. Rev. Mater.}\ }\textbf {\bibinfo {volume} {7}},\
  \bibinfo {pages} {482} (\bibinfo {year} {2022}{\natexlab{d}})}\BibitemShut
  {NoStop}%
\bibitem [{\citenamefont {Feng}\ \emph {et~al.}(2022)\citenamefont {Feng},
  \citenamefont {Zhou}, \citenamefont {$\check{\rm{S}}$mejkal}, \citenamefont
  {Wu}, \citenamefont {Zhu}, \citenamefont {Guo}, \citenamefont
  {Gonz\'alez-Hern\'andez}, \citenamefont {Wang}, \citenamefont {Yan},
  \citenamefont {Qin}, \citenamefont {Zhang}, \citenamefont {Wu}, \citenamefont
  {Chen}, \citenamefont {Meng}, \citenamefont {Liu}, \citenamefont {Xia},
  \citenamefont {Sinova}, \citenamefont {Jungwirth},\ and\ \citenamefont
  {Liu}}]{AHE-RuO2-NE2022}%
  \BibitemOpen
  \bibfield  {author} {\bibinfo {author} {\bibfnamefont {Z.}~\bibnamefont
  {Feng}}, \bibinfo {author} {\bibfnamefont {X.}~\bibnamefont {Zhou}}, \bibinfo
  {author} {\bibfnamefont {L.}~\bibnamefont {$\check{\rm{S}}$mejkal}}, \bibinfo
  {author} {\bibfnamefont {L.}~\bibnamefont {Wu}}, \bibinfo {author}
  {\bibfnamefont {Z.}~\bibnamefont {Zhu}}, \bibinfo {author} {\bibfnamefont
  {H.}~\bibnamefont {Guo}}, \bibinfo {author} {\bibfnamefont {R.}~\bibnamefont
  {Gonz\'alez-Hern\'andez}}, \bibinfo {author} {\bibfnamefont {X.}~\bibnamefont
  {Wang}}, \bibinfo {author} {\bibfnamefont {H.}~\bibnamefont {Yan}}, \bibinfo
  {author} {\bibfnamefont {P.}~\bibnamefont {Qin}}, \bibinfo {author}
  {\bibfnamefont {X.}~\bibnamefont {Zhang}}, \bibinfo {author} {\bibfnamefont
  {H.}~\bibnamefont {Wu}}, \bibinfo {author} {\bibfnamefont {H.}~\bibnamefont
  {Chen}}, \bibinfo {author} {\bibfnamefont {Z.}~\bibnamefont {Meng}}, \bibinfo
  {author} {\bibfnamefont {L.}~\bibnamefont {Liu}}, \bibinfo {author}
  {\bibfnamefont {Z.}~\bibnamefont {Xia}}, \bibinfo {author} {\bibfnamefont
  {J.}~\bibnamefont {Sinova}}, \bibinfo {author} {\bibfnamefont
  {T.}~\bibnamefont {Jungwirth}},\ and\ \bibinfo {author} {\bibfnamefont
  {Z.}~\bibnamefont {Liu}},\ }\bibfield  {title} {\bibinfo {title} {An
  anomalous hall effect in altermagnetic ruthenium dioxide},\ }\href
  {https://doi.org/10.1038/s41928-022-00866-z} {\bibfield  {journal} {\bibinfo
  {journal} {Nat. Electron.}\ }\textbf {\bibinfo {volume} {5}},\ \bibinfo
  {pages} {735} (\bibinfo {year} {2022})}\BibitemShut {NoStop}%
\bibitem [{\citenamefont {Gonzalez~Betancourt}\ \emph
  {et~al.}(2023)\citenamefont {Gonzalez~Betancourt}, \citenamefont
  {Zub\'a\ifmmode~\check{c}\else \v{c}\fi{}}, \citenamefont
  {Gonzalez-Hernandez}, \citenamefont {Geishendorf}, \citenamefont {\ifmmode
  \check{S}\else \v{S}\fi{}ob\'a\ifmmode~\check{n}\else \v{n}\fi{}},
  \citenamefont {Springholz}, \citenamefont {Olejn\'{\i}k}, \citenamefont
  {\ifmmode~\check{S}\else \v{S}\fi{}mejkal}, \citenamefont {Sinova},
  \citenamefont {Jungwirth}, \citenamefont {Goennenwein}, \citenamefont
  {Thomas}, \citenamefont {Reichlov\'a}, \citenamefont {\ifmmode~\check{Z}\else
  \v{Z}\fi{}elezn\'y},\ and\ \citenamefont {Kriegner}}]{AHE-MnTe-PRL2023}%
  \BibitemOpen
  \bibfield  {author} {\bibinfo {author} {\bibfnamefont {R.~D.}\ \bibnamefont
  {Gonzalez~Betancourt}}, \bibinfo {author} {\bibfnamefont {J.}~\bibnamefont
  {Zub\'a\ifmmode~\check{c}\else \v{c}\fi{}}}, \bibinfo {author} {\bibfnamefont
  {R.}~\bibnamefont {Gonzalez-Hernandez}}, \bibinfo {author} {\bibfnamefont
  {K.}~\bibnamefont {Geishendorf}}, \bibinfo {author} {\bibfnamefont
  {Z.}~\bibnamefont {\ifmmode \check{S}\else
  \v{S}\fi{}ob\'a\ifmmode~\check{n}\else \v{n}\fi{}}}, \bibinfo {author}
  {\bibfnamefont {G.}~\bibnamefont {Springholz}}, \bibinfo {author}
  {\bibfnamefont {K.}~\bibnamefont {Olejn\'{\i}k}}, \bibinfo {author}
  {\bibfnamefont {L.}~\bibnamefont {\ifmmode~\check{S}\else \v{S}\fi{}mejkal}},
  \bibinfo {author} {\bibfnamefont {J.}~\bibnamefont {Sinova}}, \bibinfo
  {author} {\bibfnamefont {T.}~\bibnamefont {Jungwirth}}, \bibinfo {author}
  {\bibfnamefont {S.~T.~B.}\ \bibnamefont {Goennenwein}}, \bibinfo {author}
  {\bibfnamefont {A.}~\bibnamefont {Thomas}}, \bibinfo {author} {\bibfnamefont
  {H.}~\bibnamefont {Reichlov\'a}}, \bibinfo {author} {\bibfnamefont
  {J.}~\bibnamefont {\ifmmode~\check{Z}\else \v{Z}\fi{}elezn\'y}},\ and\
  \bibinfo {author} {\bibfnamefont {D.}~\bibnamefont {Kriegner}},\ }\bibfield
  {title} {\bibinfo {title} {{Spontaneous Anomalous Hall Effect Arising from an
  Unconventional Compensated Magnetic Phase in a Semiconductor}},\ }\href
  {https://doi.org/10.1103/PhysRevLett.130.036702} {\bibfield  {journal}
  {\bibinfo  {journal} {Phys. Rev. Lett.}\ }\textbf {\bibinfo {volume} {130}},\
  \bibinfo {pages} {036702} (\bibinfo {year} {2023})}\BibitemShut {NoStop}%
\bibitem [{\citenamefont {Hou}\ \emph {et~al.}(2023)\citenamefont {Hou},
  \citenamefont {Yang}, \citenamefont {Liu}, \citenamefont {Guo},\ and\
  \citenamefont {Lu}}]{AHE-hou2023}%
  \BibitemOpen
  \bibfield  {author} {\bibinfo {author} {\bibfnamefont {X.-Y.}\ \bibnamefont
  {Hou}}, \bibinfo {author} {\bibfnamefont {H.-C.}\ \bibnamefont {Yang}},
  \bibinfo {author} {\bibfnamefont {Z.-X.}\ \bibnamefont {Liu}}, \bibinfo
  {author} {\bibfnamefont {P.-J.}\ \bibnamefont {Guo}},\ and\ \bibinfo {author}
  {\bibfnamefont {Z.-Y.}\ \bibnamefont {Lu}},\ }\bibfield  {title} {\bibinfo
  {title} {{Large intrinsic anomalous Hall effect in both Nb$_2$FeB$_2$ and
  Ta$_2$FeB$_2$ with collinear antiferromagnetism}},\ }\href
  {https://doi.org/10.1103/PhysRevB.107.L161109} {\bibfield  {journal}
  {\bibinfo  {journal} {Phys. Rev. B}\ }\textbf {\bibinfo {volume} {107}},\
  \bibinfo {pages} {L161109} (\bibinfo {year} {2023})}\BibitemShut {NoStop}%
\bibitem [{\citenamefont {Zhou}\ \emph {et~al.}(2021)\citenamefont {Zhou},
  \citenamefont {Feng}, \citenamefont {Yang}, \citenamefont {Guo},\ and\
  \citenamefont {Yao}}]{MOE-Yao2021}%
  \BibitemOpen
  \bibfield  {author} {\bibinfo {author} {\bibfnamefont {X.}~\bibnamefont
  {Zhou}}, \bibinfo {author} {\bibfnamefont {W.}~\bibnamefont {Feng}}, \bibinfo
  {author} {\bibfnamefont {X.}~\bibnamefont {Yang}}, \bibinfo {author}
  {\bibfnamefont {G.-Y.}\ \bibnamefont {Guo}},\ and\ \bibinfo {author}
  {\bibfnamefont {Y.}~\bibnamefont {Yao}},\ }\bibfield  {title} {\bibinfo
  {title} {Crystal chirality magneto-optical effects in collinear
  antiferromagnets},\ }\href {https://doi.org/10.1103/PhysRevB.104.024401}
  {\bibfield  {journal} {\bibinfo  {journal} {Phys. Rev. B}\ }\textbf {\bibinfo
  {volume} {104}},\ \bibinfo {pages} {024401} (\bibinfo {year}
  {2021})}\BibitemShut {NoStop}%
\bibitem [{\citenamefont {Zhou}\ \emph {et~al.}(2024)\citenamefont {Zhou},
  \citenamefont {Feng}, \citenamefont {Zhang}, \citenamefont
  {$\check{\rm{S}}$mejkal}, \citenamefont {Sinova}, \citenamefont {Mokrousov},\
  and\ \citenamefont {Yao}}]{CTHE-Yao2024}%
  \BibitemOpen
  \bibfield  {author} {\bibinfo {author} {\bibfnamefont {X.}~\bibnamefont
  {Zhou}}, \bibinfo {author} {\bibfnamefont {W.}~\bibnamefont {Feng}}, \bibinfo
  {author} {\bibfnamefont {R.-W.}\ \bibnamefont {Zhang}}, \bibinfo {author}
  {\bibfnamefont {L.}~\bibnamefont {$\check{\rm{S}}$mejkal}}, \bibinfo {author}
  {\bibfnamefont {J.}~\bibnamefont {Sinova}}, \bibinfo {author} {\bibfnamefont
  {Y.}~\bibnamefont {Mokrousov}},\ and\ \bibinfo {author} {\bibfnamefont
  {Y.}~\bibnamefont {Yao}},\ }\bibfield  {title} {\bibinfo {title} {{Crystal
  Thermal Transport in Altermagnetic RuO$_2$}},\ }\href
  {https://doi.org/10.1103/PhysRevLett.132.056701} {\bibfield  {journal}
  {\bibinfo  {journal} {Phys. Rev. Lett.}\ }\textbf {\bibinfo {volume} {132}},\
  \bibinfo {pages} {056701} (\bibinfo {year} {2024})}\BibitemShut {NoStop}%
\bibitem [{\citenamefont {Li}\ \emph {et~al.}(2024)\citenamefont {Li},
  \citenamefont {Liu},\ and\ \citenamefont {Liu}}]{HighoT-liu2024}%
  \BibitemOpen
  \bibfield  {author} {\bibinfo {author} {\bibfnamefont {Y.-X.}\ \bibnamefont
  {Li}}, \bibinfo {author} {\bibfnamefont {Y.}~\bibnamefont {Liu}},\ and\
  \bibinfo {author} {\bibfnamefont {C.-C.}\ \bibnamefont {Liu}},\ }\bibfield
  {title} {\bibinfo {title} {{Creation and manipulation of higher-order
  topological states by altermagnets}},\ }\href
  {https://doi.org/10.1103/PhysRevB.109.L201109} {\bibfield  {journal}
  {\bibinfo  {journal} {Phys. Rev. B}\ }\textbf {\bibinfo {volume} {109}},\
  \bibinfo {pages} {L201109} (\bibinfo {year} {2024})}\BibitemShut {NoStop}%
\bibitem [{\citenamefont {Li}\ and\ \citenamefont {Liu}(2023)}]{MCM-liu2023}%
  \BibitemOpen
  \bibfield  {author} {\bibinfo {author} {\bibfnamefont {Y.-X.}\ \bibnamefont
  {Li}}\ and\ \bibinfo {author} {\bibfnamefont {C.-C.}\ \bibnamefont {Liu}},\
  }\bibfield  {title} {\bibinfo {title} {{Majorana corner modes and tunable
  patterns in an altermagnet heterostructure}},\ }\href
  {https://doi.org/10.1103/PhysRevB.108.205410} {\bibfield  {journal} {\bibinfo
   {journal} {Phys. Rev. B}\ }\textbf {\bibinfo {volume} {108}},\ \bibinfo
  {pages} {205410} (\bibinfo {year} {2023})}\BibitemShut {NoStop}%
\bibitem [{\citenamefont {Guo}\ \emph {et~al.}(2023{\natexlab{b}})\citenamefont
  {Guo}, \citenamefont {Gu}, \citenamefont {Gao},\ and\ \citenamefont
  {Lu}}]{LiFe2F6-guo2023}%
  \BibitemOpen
  \bibfield  {author} {\bibinfo {author} {\bibfnamefont {P.-J.}\ \bibnamefont
  {Guo}}, \bibinfo {author} {\bibfnamefont {Y.}~\bibnamefont {Gu}}, \bibinfo
  {author} {\bibfnamefont {Z.-F.}\ \bibnamefont {Gao}},\ and\ \bibinfo {author}
  {\bibfnamefont {Z.-Y.}\ \bibnamefont {Lu}},\ }\bibfield  {title} {\bibinfo
  {title} {{Altermagnetic ferroelectric LiFe$_2$F$_6$ and spin-triplet
  excitonic insulator phase}},\ }\href@noop {} {\  (\bibinfo {year}
  {2023}{\natexlab{b}})},\ \Eprint {https://arxiv.org/abs/2312.13911}
  {arXiv:2312.13911 [cond-mat.mtrl-sci]} \BibitemShut {NoStop}%
\bibitem [{\citenamefont {Qu}\ \emph {et~al.}(2024)\citenamefont {Qu},
  \citenamefont {Gao}, \citenamefont {Sun}, \citenamefont {Liu}, \citenamefont
  {Guo},\ and\ \citenamefont {Lu}}]{NiF3-qu2024}%
  \BibitemOpen
  \bibfield  {author} {\bibinfo {author} {\bibfnamefont {S.}~\bibnamefont
  {Qu}}, \bibinfo {author} {\bibfnamefont {Z.-F.}\ \bibnamefont {Gao}},
  \bibinfo {author} {\bibfnamefont {H.}~\bibnamefont {Sun}}, \bibinfo {author}
  {\bibfnamefont {K.}~\bibnamefont {Liu}}, \bibinfo {author} {\bibfnamefont
  {P.-J.}\ \bibnamefont {Guo}},\ and\ \bibinfo {author} {\bibfnamefont {Z.-Y.}\
  \bibnamefont {Lu}},\ }\bibfield  {title} {\bibinfo {title} {{Extremely strong
  spin-orbit coupling effect in light element altermagnetic materials}},\
  }\href@noop {} {\  (\bibinfo {year} {2024})},\ \Eprint
  {https://arxiv.org/abs/2401.11065} {arXiv:2401.11065 [cond-mat.mtrl-sci]}
  \BibitemShut {NoStop}%
\bibitem [{\citenamefont {Tan}\ \emph {et~al.}(2024{\natexlab{a}})\citenamefont
  {Tan}, \citenamefont {Gao}, \citenamefont {Yang}, \citenamefont {Liu},
  \citenamefont {Guo},\ and\ \citenamefont {Lu}}]{BWS}%
  \BibitemOpen
  \bibfield  {author} {\bibinfo {author} {\bibfnamefont {C.-Y.}\ \bibnamefont
  {Tan}}, \bibinfo {author} {\bibfnamefont {Z.-F.}\ \bibnamefont {Gao}},
  \bibinfo {author} {\bibfnamefont {H.-C.}\ \bibnamefont {Yang}}, \bibinfo
  {author} {\bibfnamefont {K.}~\bibnamefont {Liu}}, \bibinfo {author}
  {\bibfnamefont {P.-J.}\ \bibnamefont {Guo}},\ and\ \bibinfo {author}
  {\bibfnamefont {Z.-Y.}\ \bibnamefont {Lu}},\ }\bibfield  {title} {\bibinfo
  {title} {Bipolarized weyl semimetals and quantum crystal valley hall effect
  in two-dimensional altermagnetic materials},\ }\href@noop {} {\  (\bibinfo
  {year} {2024}{\natexlab{a}})},\ \Eprint {https://arxiv.org/abs/2406.16603}
  {arXiv:2406.16603 [cond-mat.mtrl-sci]} \BibitemShut {NoStop}%
\bibitem [{\citenamefont {Tan}\ \emph {et~al.}(2024{\natexlab{b}})\citenamefont
  {Tan}, \citenamefont {Gao}, \citenamefont {Yang}, \citenamefont {Liu},
  \citenamefont {Liu}, \citenamefont {Guo},\ and\ \citenamefont {Lu}}]{CVHE}%
  \BibitemOpen
  \bibfield  {author} {\bibinfo {author} {\bibfnamefont {C.-Y.}\ \bibnamefont
  {Tan}}, \bibinfo {author} {\bibfnamefont {Z.-F.}\ \bibnamefont {Gao}},
  \bibinfo {author} {\bibfnamefont {H.-C.}\ \bibnamefont {Yang}}, \bibinfo
  {author} {\bibfnamefont {Z.-X.}\ \bibnamefont {Liu}}, \bibinfo {author}
  {\bibfnamefont {K.}~\bibnamefont {Liu}}, \bibinfo {author} {\bibfnamefont
  {P.-J.}\ \bibnamefont {Guo}},\ and\ \bibinfo {author} {\bibfnamefont {Z.-Y.}\
  \bibnamefont {Lu}},\ }\bibfield  {title} {\bibinfo {title} {Crystal valley
  hall effect},\ }\href@noop {} {\  (\bibinfo {year} {2024}{\natexlab{b}})},\
  \Eprint {https://arxiv.org/abs/2410.00073} {arXiv:2410.00073
  [cond-mat.mtrl-sci]} \BibitemShut {NoStop}%
\bibitem [{\citenamefont {Xiao}\ \emph {et~al.}(2024)\citenamefont {Xiao},
  \citenamefont {Zhao}, \citenamefont {Li}, \citenamefont {Shindou},\ and\
  \citenamefont {Song}}]{Song-PRX}%
  \BibitemOpen
  \bibfield  {author} {\bibinfo {author} {\bibfnamefont {Z.}~\bibnamefont
  {Xiao}}, \bibinfo {author} {\bibfnamefont {J.}~\bibnamefont {Zhao}}, \bibinfo
  {author} {\bibfnamefont {Y.}~\bibnamefont {Li}}, \bibinfo {author}
  {\bibfnamefont {R.}~\bibnamefont {Shindou}},\ and\ \bibinfo {author}
  {\bibfnamefont {Z.-D.}\ \bibnamefont {Song}},\ }\bibfield  {title} {\bibinfo
  {title} {Spin space groups: Full classification and applications},\ }\href
  {https://doi.org/10.1103/PhysRevX.14.031037} {\bibfield  {journal} {\bibinfo
  {journal} {Phys. Rev. X}\ }\textbf {\bibinfo {volume} {14}},\ \bibinfo
  {pages} {031037} (\bibinfo {year} {2024})}\BibitemShut {NoStop}%
\bibitem [{\citenamefont {Chen}\ \emph {et~al.}(2024)\citenamefont {Chen},
  \citenamefont {Ren}, \citenamefont {Zhu}, \citenamefont {Yu}, \citenamefont
  {Zhang}, \citenamefont {Liu}, \citenamefont {Li}, \citenamefont {Liu},
  \citenamefont {Li},\ and\ \citenamefont {Liu}}]{Liu-PRX}%
  \BibitemOpen
  \bibfield  {author} {\bibinfo {author} {\bibfnamefont {X.}~\bibnamefont
  {Chen}}, \bibinfo {author} {\bibfnamefont {J.}~\bibnamefont {Ren}}, \bibinfo
  {author} {\bibfnamefont {Y.}~\bibnamefont {Zhu}}, \bibinfo {author}
  {\bibfnamefont {Y.}~\bibnamefont {Yu}}, \bibinfo {author} {\bibfnamefont
  {A.}~\bibnamefont {Zhang}}, \bibinfo {author} {\bibfnamefont
  {P.}~\bibnamefont {Liu}}, \bibinfo {author} {\bibfnamefont {J.}~\bibnamefont
  {Li}}, \bibinfo {author} {\bibfnamefont {Y.}~\bibnamefont {Liu}}, \bibinfo
  {author} {\bibfnamefont {C.}~\bibnamefont {Li}},\ and\ \bibinfo {author}
  {\bibfnamefont {Q.}~\bibnamefont {Liu}},\ }\bibfield  {title} {\bibinfo
  {title} {Enumeration and representation theory of spin space groups},\ }\href
  {https://doi.org/10.1103/PhysRevX.14.031038} {\bibfield  {journal} {\bibinfo
  {journal} {Phys. Rev. X}\ }\textbf {\bibinfo {volume} {14}},\ \bibinfo
  {pages} {031038} (\bibinfo {year} {2024})}\BibitemShut {NoStop}%
\bibitem [{\citenamefont {Gao}\ \emph {et~al.}(2023)\citenamefont {Gao},
  \citenamefont {Qu}, \citenamefont {Zeng}, \citenamefont {Liu}, \citenamefont
  {Wen}, \citenamefont {Sun}, \citenamefont {Guo},\ and\ \citenamefont
  {Lu}}]{Gao-AI}%
  \BibitemOpen
  \bibfield  {author} {\bibinfo {author} {\bibfnamefont {Z.-F.}\ \bibnamefont
  {Gao}}, \bibinfo {author} {\bibfnamefont {S.}~\bibnamefont {Qu}}, \bibinfo
  {author} {\bibfnamefont {B.-C.}\ \bibnamefont {Zeng}}, \bibinfo {author}
  {\bibfnamefont {Y.}~\bibnamefont {Liu}}, \bibinfo {author} {\bibfnamefont
  {J.-R.}\ \bibnamefont {Wen}}, \bibinfo {author} {\bibfnamefont
  {H.}~\bibnamefont {Sun}}, \bibinfo {author} {\bibfnamefont {P.-J.}\
  \bibnamefont {Guo}},\ and\ \bibinfo {author} {\bibfnamefont {Z.-Y.}\
  \bibnamefont {Lu}},\ }\href@noop {} {\bibinfo {title} {Ai-accelerated
  discovery of altermagnetic materials}} (\bibinfo {year} {2023}),\ \Eprint
  {https://arxiv.org/abs/2311.04418} {arXiv:2311.04418 [cond-mat.mtrl-sci]}
  \BibitemShut {NoStop}%
\bibitem [{\citenamefont {Luttinger}\ and\ \citenamefont
  {Ward}(1960)}]{Luttinger-I}%
  \BibitemOpen
  \bibfield  {author} {\bibinfo {author} {\bibfnamefont {J.~M.}\ \bibnamefont
  {Luttinger}}\ and\ \bibinfo {author} {\bibfnamefont {J.~C.}\ \bibnamefont
  {Ward}},\ }\bibfield  {title} {\bibinfo {title} {Ground-state energy of a
  many-fermion system},\ }\href {https://doi.org/10.1103/PhysRev.118.1417}
  {\bibfield  {journal} {\bibinfo  {journal} {Phys. Rev.}\ }\textbf {\bibinfo
  {volume} {118}},\ \bibinfo {pages} {1417} (\bibinfo {year}
  {1960})}\BibitemShut {NoStop}%
\bibitem [{\citenamefont {Luttinger}(1960)}]{Luttinger-II}%
  \BibitemOpen
  \bibfield  {author} {\bibinfo {author} {\bibfnamefont {J.~M.}\ \bibnamefont
  {Luttinger}},\ }\bibfield  {title} {\bibinfo {title} {Fermi surface and some
  simple equilibrium properties of a system of interacting fermions},\ }\href
  {https://doi.org/10.1103/PhysRev.119.1153} {\bibfield  {journal} {\bibinfo
  {journal} {Phys. Rev.}\ }\textbf {\bibinfo {volume} {119}},\ \bibinfo {pages}
  {1153} (\bibinfo {year} {1960})}\BibitemShut {NoStop}%
\bibitem [{\citenamefont {Mazin}(2022{\natexlab{b}})}]{MazinPRX2022}%
  \BibitemOpen
  \bibfield  {author} {\bibinfo {author} {\bibfnamefont {I.}~\bibnamefont
  {Mazin}},\ }\bibfield  {title} {\bibinfo {title} {Editorial:
  Altermagnetism---a new punch line of fundamental magnetism},\ }\href
  {https://doi.org/10.1103/PhysRevX.12.040002} {\bibfield  {journal} {\bibinfo
  {journal} {Phys. Rev. X}\ }\textbf {\bibinfo {volume} {12}},\ \bibinfo
  {pages} {040002} (\bibinfo {year} {2022}{\natexlab{b}})}\BibitemShut
  {NoStop}%
\bibitem [{\citenamefont {Kresse}\ and\ \citenamefont
  {Furthm\..uller}(1996)}]{Vasp-1996}%
  \BibitemOpen
  \bibfield  {author} {\bibinfo {author} {\bibfnamefont {G.}~\bibnamefont
  {Kresse}}\ and\ \bibinfo {author} {\bibfnamefont {J.}~\bibnamefont
  {Furthm\..uller}},\ }\bibfield  {title} {\bibinfo {title} {{Efficiency of
  ab-initio total energy calculations for metals and semiconductors using a
  plane-wave basis set}},\ }\href
  {https://doi.org/https://doi.org/10.1016/0927-0256(96)00008-0} {\bibfield
  {journal} {\bibinfo  {journal} {Comput. Mater. Sci.}\ }\textbf {\bibinfo
  {volume} {6}},\ \bibinfo {pages} {15} (\bibinfo {year} {1996})}\BibitemShut
  {NoStop}%
\bibitem [{\citenamefont {Bl\"ochl}(1994)}]{PAW-1994}%
  \BibitemOpen
  \bibfield  {author} {\bibinfo {author} {\bibfnamefont {P.~E.}\ \bibnamefont
  {Bl\"ochl}},\ }\bibfield  {title} {\bibinfo {title} {{Projector
  augmented-wave method}},\ }\href {https://doi.org/10.1103/PhysRevB.50.17953}
  {\bibfield  {journal} {\bibinfo  {journal} {Phys. Rev. B}\ }\textbf {\bibinfo
  {volume} {50}},\ \bibinfo {pages} {17953} (\bibinfo {year}
  {1994})}\BibitemShut {NoStop}%
\bibitem [{\citenamefont {Perdew}\ \emph {et~al.}(1996)\citenamefont {Perdew},
  \citenamefont {Burke},\ and\ \citenamefont {Ernzerhof}}]{GGA-1996}%
  \BibitemOpen
  \bibfield  {author} {\bibinfo {author} {\bibfnamefont {J.~P.}\ \bibnamefont
  {Perdew}}, \bibinfo {author} {\bibfnamefont {K.}~\bibnamefont {Burke}},\ and\
  \bibinfo {author} {\bibfnamefont {M.}~\bibnamefont {Ernzerhof}},\ }\bibfield
  {title} {\bibinfo {title} {{Generalized Gradient Approximation Made
  Simple}},\ }\href {https://doi.org/10.1103/PhysRevLett.77.3865} {\bibfield
  {journal} {\bibinfo  {journal} {Phys. Rev. Lett.}\ }\textbf {\bibinfo
  {volume} {77}},\ \bibinfo {pages} {3865} (\bibinfo {year}
  {1996})}\BibitemShut {NoStop}%
\bibitem [{\citenamefont {Anisimov}\ \emph {et~al.}(1991)\citenamefont
  {Anisimov}, \citenamefont {Zaanen},\ and\ \citenamefont {Andersen}}]{LDAU1}%
  \BibitemOpen
  \bibfield  {author} {\bibinfo {author} {\bibfnamefont {V.~I.}\ \bibnamefont
  {Anisimov}}, \bibinfo {author} {\bibfnamefont {J.}~\bibnamefont {Zaanen}},\
  and\ \bibinfo {author} {\bibfnamefont {O.~K.}\ \bibnamefont {Andersen}},\
  }\bibfield  {title} {\bibinfo {title} {Band theory and mott insulators:
  Hubbard {U} instead of stoner {I}},\ }\href
  {https://doi.org/10.1103/PhysRevB.44.943} {\bibfield  {journal} {\bibinfo
  {journal} {Phys. Rev. B}\ }\textbf {\bibinfo {volume} {44}},\ \bibinfo
  {pages} {943} (\bibinfo {year} {1991})}\BibitemShut {NoStop}%
\bibitem [{\citenamefont {Dudarev}\ \emph {et~al.}(1998)\citenamefont
  {Dudarev}, \citenamefont {Botton}, \citenamefont {Savrasov}, \citenamefont
  {Humphreys},\ and\ \citenamefont {Sutton}}]{LDAU2}%
  \BibitemOpen
  \bibfield  {author} {\bibinfo {author} {\bibfnamefont {S.~L.}\ \bibnamefont
  {Dudarev}}, \bibinfo {author} {\bibfnamefont {G.~A.}\ \bibnamefont {Botton}},
  \bibinfo {author} {\bibfnamefont {S.~Y.}\ \bibnamefont {Savrasov}}, \bibinfo
  {author} {\bibfnamefont {C.~J.}\ \bibnamefont {Humphreys}},\ and\ \bibinfo
  {author} {\bibfnamefont {A.~P.}\ \bibnamefont {Sutton}},\ }\bibfield  {title}
  {\bibinfo {title} {Electron-energy-loss spectra and the structural stability
  of nickel oxide: An {LSDA+U} study},\ }\href
  {https://doi.org/10.1103/PhysRevB.57.1505} {\bibfield  {journal} {\bibinfo
  {journal} {Phys. Rev. B}\ }\textbf {\bibinfo {volume} {57}},\ \bibinfo
  {pages} {1505} (\bibinfo {year} {1998})}\BibitemShut {NoStop}%
\bibitem [{\citenamefont {Liu}\ \emph {et~al.}(2019)\citenamefont {Liu},
  \citenamefont {Ren}, \citenamefont {Xie}, \citenamefont {Cheng},
  \citenamefont {Liu}, \citenamefont {An}, \citenamefont {Qin},\ and\
  \citenamefont {Hu}}]{Mcsolver}%
  \BibitemOpen
  \bibfield  {author} {\bibinfo {author} {\bibfnamefont {L.}~\bibnamefont
  {Liu}}, \bibinfo {author} {\bibfnamefont {X.}~\bibnamefont {Ren}}, \bibinfo
  {author} {\bibfnamefont {J.~H.}\ \bibnamefont {Xie}}, \bibinfo {author}
  {\bibfnamefont {B.}~\bibnamefont {Cheng}}, \bibinfo {author} {\bibfnamefont
  {W.~K.}\ \bibnamefont {Liu}}, \bibinfo {author} {\bibfnamefont {T.~Y.}\
  \bibnamefont {An}}, \bibinfo {author} {\bibfnamefont {H.~W.}\ \bibnamefont
  {Qin}},\ and\ \bibinfo {author} {\bibfnamefont {J.~F.}\ \bibnamefont {Hu}},\
  }\bibfield  {title} {\bibinfo {title} {Magnetic switches via electric field
  in bn nanoribbons},\ }\href {https://doi.org/10.1016/j.apsusc.2019.02.203}
  {\bibfield  {journal} {\bibinfo  {journal} {Appl. Surf. Sci.}\ }\textbf
  {\bibinfo {volume} {480}},\ \bibinfo {pages} {300} (\bibinfo {year}
  {2019})}\BibitemShut {NoStop}%
\bibitem [{\citenamefont {Goodenough}(1963)}]{GKA}%
  \BibitemOpen
  \bibfield  {author} {\bibinfo {author} {\bibfnamefont {J.}~\bibnamefont
  {Goodenough}},\ }\bibfield  {title} {\bibinfo {title} {Magnetism and the
  chemical bond},\ }\href@noop {} {\bibfield  {journal} {\bibinfo  {journal}
  {Interscience Publishers}\ } (\bibinfo {year} {1963})}\BibitemShut {NoStop}%
\end{thebibliography}%

\end{document}